\documentclass[authoryear,preprint,review,12p]{elsarticle}
\usepackage{amssymb,amsmath, morefloats, graphicx,float,threeparttable,rotating,longtable,array,pifont, geometry,fleqn, txfonts, hyperref,url}
\journal{Icarus }
\begin{document}

\begin{frontmatter}

\title{Absolute magnitudes and slope parameters for 250,000 asteroids observed by Pan-STARRS PS1 - preliminary results.}

\author[fmfi,ifa]{P. Vere\v{s}}
\ead{veres@fmph.uniba.sk}

\author[ifa]{R. Jedicke}
\author[queens]{A. Fitzsimmons}
\author[ifa]{L. Denneau}
 \author[fin,fingeo]{M. Granvik}
  \author[ifa,fra]{B. Bolin}
  \author[ifa]{S. Chastel}
  \author[ifa]{R.~J. Wainscoat}
  \author[gmto]{W.~S. Burgett}
   \author[ifa]{K.~C. Chambers}
 \author[ifa]{H. Flewelling}
 \author[ifa]{N. Kaiser}
 \author[ifa]{E~.A. Magnier}
\author[ifa]{J.~S. Morgan}
\author[ifa]{Paul A. Price}
\author[ifa]{J.~L.Tonry}
\author[ifa]{C. Waters}

\address[fmfi]{Faculty of Mathematics, Physics and Informatics, Comenius University in Bratislava, Mlynsk\'{a} Dolina F1, 84248 Bratislava, Slovakia}
\address[ifa]{Institute for Astronomy, University of Hawaii at Manoa, 2680 Woodlawn Drive, Honolulu, HI 96822, USA}
\address[queens]{Queen's University Belfast, Belfast BT7 1NN, Northern Ireland, UK}
\address[fin]{Department of Physics, P.O. Box 64, 00014 University of Helsinki, Finland}
\address[fingeo]{Finnish Geospatial Research Institute, P.O. Box 15, 02430 Masala, Finland}
\address[fra]{UNS-CNRS-Observatoire de la C\^{o}te d`Azur, BP 4229, 06304 Nice Cedex 4, France}
\address[gmto]{GMTO Corp., 251 S. Lake Ave., Suite 300, Pasadena, CA  91101, USA}

\begin{abstract}

We present the results of a Monte Carlo technique to calculate the
absolute magnitudes ($H$) and slope parameters ($G$) of $\sim240,000$
asteroids observed by the Pan-STARRS1 telescope during the first 15 months
of its 3-year all-sky survey mission. The system's exquisite
photometry with photometric errors $\le0.04$\,mags, and well-defined
filter and photometric system, allowed us to derive accurate $H$ and $G$ even
with a limited number of observations and restricted range in phase
angles.  Our Monte Carlo method simulates each asteroid's rotation period,
amplitude and color to derive the most-likely $H$ and $G$, but its major advantage is in estimating realistic statistical+systematic uncertainties and errors on each
parameter.  The method was confirmed by comparison with the
well-established and accurate results for about 500 asteroids provided
by \citet{Pra12} and then applied to determining $H$ and $G$ for the
Pan-STARRS1 asteroids using both the \citet{Mui10} and \citet{Bow89} phase
functions.

\end{abstract}


\end{frontmatter}

\section{Introduction}

Asteroid diameters are critical to understanding their dynamical and
morphological evolution, potential as spacecraft targets, impact
threat, and much more, yet most asteroid diameters are uncertain by
$>50$\% because of the difficulties involved in calculating diameter
from apparent brightness.  The problem is that an asteroid's apparent
brightness is a complicated function of the observing geometry, their
irregular shapes, rotation phase, albedo, lack of atmosphere, and
their rough, regolith-covered surfaces.  Most of these data are
unknown for most asteroids. The issue has been further confused
because catalogued apparent magnitudes for individual asteroids may have been
reported by numerous observers and observatories over many years
(even decades) in a variety of photometric systems with varying
concern for ensuring accuracy and precision.  This work describes our process for calculating asteroid absolute magnitudes (from which diameter is
calculated) and their statistical and systematic uncertainties for hundreds of thousands of asteroids using sparse but accurate and
precise data from a single observatory, the Pan-STARRS1 facility on Maui,
HI, USA.  Our technique is suited to estimating absolute magnitudes when the phase curve coverage is even more sparse than those obtained by the Palomar Transient Factory \citep{Law09}.

An asteroid's absolute magnitude, $H$, is the apparent Johnson V-band magnitude, $m$, it would have if observed from the Sun at a distance of $1\,au$ (i.e. observed at zero phase angle and $1\,au$ distance).  Accurate
measurements of $H$ as a function of time, together with infrared, polarimetric and radiometric observations, can provide crucial information about an
asteroid's size and shape, geometric albedo, surface properties
and spin characteristics.

In 1985 the International Astronomical Union (IAU) adopted the
two-parameter phase function developed by \citet[][hereafter B89]{Bow89}, $\Phi_{B}(\alpha;H_B,G_B)$, describing the behavior of the apparent
magnitude:
\begin{equation}
\centering
m(r,\Delta;H_B,G_B) = 5\log(r\Delta) + \Phi_{B}(\alpha;H_B,G_B)
\label{eq.Bowell}
\end{equation}

\noindent where $\Delta$ represents the topocentric distance, $r$ the
heliocentric distance, and $\alpha(r,\Delta)$ is the phase angle, the angle between the Earth and Sun as observed from the asteroid. We denote absolute magnitude in the B89 system as $H_B$ with a corresponding slope parameter, $G_B$,
that depends in a non-analytical manner on (at least) an asteroid's albedo
and spectral type \citep[B89;][]{Lag90}.  The slope parameter determines how strongly the apparent brightness of an asteroid depends on the phase angle and accounts for the properties of scattered light on the asteroid's surfaces.  $G_B$ has an average value of $\sim0.15$ (B89) for the
most numerous S and C-class main belt asteroid taxonomies.  An accurate determination of both $H_B$ and
$G_B$ requires a wide and dense time coverage of the object's apparent magnitude.  Therefore, it is not surprising that only a few tens of slope parameters
were measured before the advent of dedicated CCD asteroid surveys.

The B89 phase function was very successful, but observations in the
past twenty years have shown it can not reproduce the opposition
brightening of E-type asteroids, the linear phase curve of the F-type
asteroids, and fails to accurately predict the apparent brightness of
asteroids at small phase angles. To address these issues
\citet[][hereafter M10]{Mui10} introduced an alternative phase
function, $\phi_{M}$, with two slope parameters, $G_{1}$ and $G_{2}$
 that uses cubic
splines to more accurately describe the behavior of the apparent
magnitude.  An alternative M10 formulation with a single slope parameter, $G_{12}$ that is denoted in our work as $G_M$, can be used when the data are not sufficient to derive the values of the two-parameter formulation i.e. $m = 5\log(r\Delta) + \Phi_{M}(\alpha;H_M,G_M)$.
Their phase function was constructed such that $H_M \sim H_B$ and the
average asteroid would have a slope parameter of $G_M\sim0.5$.  This
form of the phase function can provide better apparent magnitude
predictions but derivation of $H_M$ and $G_M$ still requires extensive
light curve coverage and well-calibrated observational data
\citep{Osk12}. The IAU adopted the M10 $(H,G1,G2)$ system as the new
              photometric system for asteroids in 2012.

In the remainder of this work we use $H$ and $G$ to represent 
`generic' absolute magnitudes and slope parameters respectively, and use
the subscripts $B$ and $M$ on each parameter when referring to the
values calculated using the B89 and M10 phase functions
respectively.  We implemented both functions to facilitate comparison with 1)
past work that used the B89 parameterization and 2) future
work that will use the now-standard M10 implementation.
When we use $G_M$ we specifically mean the M10 $G_{12}$
parameter.

The accuracy of most reported absolute magnitudes is poor due to the
lack of good photometry and limited phase curve
coverage. \citet[e.g.][]{Jur02} first reported a systematic error of
about $0.4$\,mags in the MPC's absolute magnitudes which the MPC (and
others) now attempt to address with observatory-dependent corrections
to the reported apparent magnitudes.

The determination of $G$ has traditionally been even more of a
challenge --- they are so difficult to measure that they have only
been calculated for $\ll0.1\%$ of asteroids and, even then, the
uncertainty is usually large \citep{Pra12}.  An accurate measurement
requires dense coverage of the phase curve and observations at
different viewing aspects on the asteroid i.e. sub-solar positions.  The vast majority of asteroids have no measured slope parameter so the
average values of $G_B=0.15$ or $G_M=0.5$ are used. This
assumption translates into a systematic error in an individual asteroid's $H$ and $G$, and large uncertainty on the distribution of the parameters in the population. The problem is particularly acute for objects that have
been observed only at large phase angles e.g. resonant objects
like 3753 Cruithne \citep{Marcos13,Wiegert97}, and objects that
orbit the Sun entirely within Earth's orbit \citep{Zavodny08} for
which absolute magnitudes might be in error by up to about 1\,mags.

In summary, the problems with our current knowledge of asteroid absolute magnitudes and slope parameters are due to:
\begin{enumerate}

\item Reporting observations to the Minor Planet Center (MPC) in non-standard filters and/or without accurate calibration.

\item Not performing the color transformation from the filter used for an
  observation to the Johnson V-band for an asteroid's (usually
  unknown) color.

\item The lack of information about the photometric uncertainty on each observation reported to the MPC so that it must be statistically `back-calculated' for each observatory (or observer) from historical observations.

\item The MPC database storing photometric values with only $0.1$\,mags precision.

\item Assuming that $G_B=0.15$ for all asteroids that
  do not have a reported value for the slope parameter.
  
\item The accepted `standard' average slope parameter of $G_B=0.15$
  for S and C class asteroids being different from the actual value of
  $G_B=0.20$ \citep{Pra12}.

\item Sparse observations (in time).  The lack of information about their rotation amplitudes induces an error and uncertainty in $H$.

\item Selection effects \citep{Jedicke2002} that bias the discovery of
  asteroids towards their rotation amplitude maxima which induce a
  systematic error in their derived $H$.

\item Most of the effort in deriving $H$ and $G$ focuses on their statistical uncertainties when the systematic uncertainties dominate. 

\end{enumerate}

In this work we address each of these issues and derive the
$(H_B,G_B)$ and $(H_M,G_M)$ parameters for known asteroids in the
inner solar system out to, and including, Jupiter's Trojan asteroids.
All the data were acquired by a single wide-field survey,
Pan-STARRS1 \citep{Kai10}, in standard filters with measured
transformations to an accepted photometric system yielding photometric
uncertainties that are typically about an order of magnitude smaller
than earlier surveys.  We use a Monte Carlo technique to
measure the systematic errors introduced by filter transformations for
unknown spectral types, unknown $G$, and the unknown asteroid spin and
amplitude.

\section{Pan-STARRS1 asteroids.}
\label{s.PS1Asteroids}

The Panoramic Survey Telescope and Rapid Response System's prototype telescope
\citep[Pan-STARRS1;][]{Kai10} was operated by the PS1 Science Consortium
during the time period in which the data used in this study was
acquired.  The telescope has a 1.4~gigapixel camera \citep{Ton09} and
$1.8\,meter$ f/4 Ritchey-Chretien optical assembly and has been
surveying the sky since the second half of 2011. Although the
scientific scope of the survey is wide --- including the solar system, exoplanets, brown dwarfs, stellar astronomy, galaxies,
cosmology, etc. --- most of the data products are suitable for asteroid
science.  About 5\% of the survey time was dedicated to the `Solar
System' (SS) survey (more accurately a survey for near-Earth objects,
NEO) through the end of 2012, was increased to about 11\% from then
till 2014 March 31, and the system is now 100\% dedicated to NEO
surveying.

\ Pan-STARRS1 surveys in six broadband filters, four of which were
designed to be similar to the Sloan Digital Sky Survey photometric
system \citep[SDSS;][]{Sloan}. Most of the observing time was devoted
to the 3$\pi$ survey of the sky north of $-30$\,arcdeg declination
for which each field was observed up to $20\times$/year in each of 5 filters ---
$g_{\rm P1}$, $r_{\rm P1}$, $i_{\rm P1}$, $z_{\rm P1}$ and $y_{\rm P1}$. In the 3$\pi$ survey the same field
is observed 2 or 4 times on a single night in 30-40\,second exposures
obtained within about an hour.  The dedicated solar system survey used
only the wide-band $w_{\rm P1}$ filter that is roughly equivalent to
$g_{\rm P1}+r_{\rm P1}+i_{\rm P1}$ with 45\,second exposures and a cadence of
$\sim20\min$ to image the same field $4\times$/night. The SS survey typically
included fields within about 30\,arcdeg of opposition or at small solar elongations ranging from 60\,arcdeg to 90\,arcdeg of the Sun.

Image processing was performed automatically and almost in real time
by the Image Processing Pipeline \citep[IPP;][]{Mag06}.  Transient
objects were identified after `difference imaging' \citep{Lup07} in
which two consecutive images were convolved and subtracted to identify
moving, or stationary but variable, targets. The photometric
calibration until May 2012 was based on combined fluxes of bright stars from Tycho,
USNO-B and 2MASS catalogs.  Since that time the entire
northern sky has been imaged by Pan-STARRS1 in all 5 filters allowing the
development and use of the Pan-STARRS1 star catalog with `ubercalibrated'
magnitudes and zero points providing photometric uncertainties of
$\sim1\%$ \citep{Sch12, Mag13}.

Moving transient detections are identified and linked into tracklets
by the Moving Object Processing System \citep[MOPS;][]{Den13} and
tracklets are associated with known asteroids by
known server\ \citep{Mil08}. As of May 2015 the Pan-STARRS1 MOPS team
 has submitted $\sim16,700,000$ positions and magnitudes of
575,000 known asteroids to the MPC representing 85\% of all numbered
asteroids. During the same time period the system discovered $\sim$41,000 asteroids, among
them about 850 NEOs and 46 comets, and reported about 2,500,000
detections of unknown asteroids to the MPC. About $\sim42$\% of the detections were in the $w_{\rm P1}$ filter
acquired during the solar system survey while only about 9\% were in
the $y_{\rm P1}$ and $z_{\rm P1}$ bands.

\begin{figure}[!ht]
\center
\includegraphics{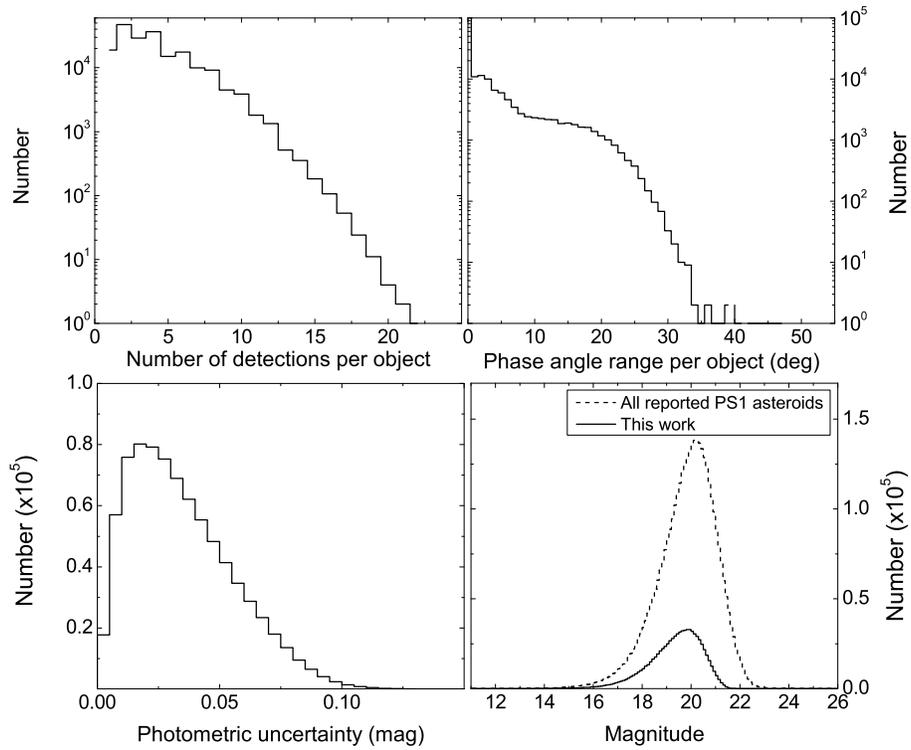}
\caption{Characteristics of PS1 asteroid detections used in this
  work. (clockwise from top left) number of detections per object,
  phase angle range per object, apparent V-magnitudes, and
  photometric uncertainties per detection.}
\label{fig.PS1asteroids}
\end{figure}

To ensure a consistent data set of high quality photometry
(Fig.~\ref{fig.PS1asteroids}) we restricted the detections used in
this study to known asteroids in the inner solar system (out to
and including Jupiter's Trojans) with multi-opposition orbits acquired during a
sub-set of the 3$\pi$ and solar system surveys between February 2011 and May
2012 (see Table \ref{Tab.Filters}) The detections were selected from
the IPP's calibrated chip-stage PSF-fit photometry \citep{Sch12} and
were required to be unsaturated, with $sn>$5, and not blended with
stars or image artifacts.  The Pan-STARRS1 IPP never implemented the
capability of fitting trailed asteroid detections, so we restricted our
data sample to asteroids that trailed by less than 5 pixels
during the exposure, equivalent to the typical PSF-width of
$\sim1\,arcsec$.  This limited the maximum rate of motion of the
asteroids to about $0.75$~deg/day, excluding most NEOs and even
fast-moving asteroids like Hungarias and Phocaeas on the inner edge of
the main belt.  Our strict criteria resulted in a set of more than one
million detections of approximately 240,000 asteroids

\begin{table}[htdp]
\caption{Percentage of Pan-STARRS1 asteroid detections in each filter in the time period from February 2011 to May 2012 (values do not add to 100\% due to rounding).}
\begin{center}
\begin{tabular}{c|cccccc}
\hline
Band &$g_{\rm P1}$ &$r_{\rm P1}$ &    $i_{\rm P1}$ & $y_{\rm P1}$ &$z_{\rm P1}$  &  $w_{\rm P1}$  \\
\hline
Fraction (\%) & 18 & 20 & 17 & 2.2 & 6.2 & 36\\
\hline
\end{tabular}
\end{center}
\label{Tab.Filters}
\end{table}

\

Despite the enormous number of asteroid detections there are only
about 10 detections/asteroid and each object is observed on average on
only $\sim3$ different nights over a phase angle range spanning about
7\,arcdeg (Fig.~\ref{fig.PS1asteroids}). Therefore, the survey
pattern does not typically allow the determination of an asteroid's
spectral type, rotation amplitude or period.  The detections have a
mean$\pm$RMS photometric uncertainty of $0.04\pm0.02$\,mags and average$\pm$RMS
visual magnitude of $19.8\pm1.2$\,mags.  The photometric
uncertainty mode is $\sim0.02$\,mags corresponding to $sm\sim50$ detections.
This surprisingly high value is due to our selection criteria: the
multi-opposition objects were identified in earlier surveys with
smaller telescopes so they are typically brighter when observed with
Pan-STARRS1.  Note that only $\sim1$\% of the detections in our data
sub-set have a photometric uncertainty greater than the $0.1$\,mags
precision provided by the MPC.

\section{Method}
\label{sec:Method}

This work introduces a Monte Carlo technique to determine $H$ (and $G$
when possible) and its statistical+systematic uncertainty based on the generation of synthetic asteroids (clones) that are each consistent with the known asteroid. The clones explore the phase space of light curve rotation
amplitudes, periods, colors and slope parameter in an attempt to replicate the
observed apparent magnitudes. Each clone's observations are evaluated individually in
the fitting process to derive $H$ and $G$ in the same manner as the
actual observations so that the distribution of values for each object's
clones provide a measure of the systematic errors in the values.

\subsection{Step 1:  Initial fit for $H$ and $G$}
\label{sss.InitialFit}

The first step is essentially identical to the typical technique for
calculating $H$ and $G$: we fit the apparent V-band magnitude to the
B89 and M10 phase functions using the IDL procedure {\tt
  mpfit2dfun}\footnote{\,Markwardt IDL library,
  \url{http://wwﬁw.physics.wisc.edu/~craigm/idl}} that employs the
Levenberg-Marquardt least-squares fitting technique
\citep{Lev44,Mar63} to minﬁimize the variance between the detections'
apparent magnitudes and the values predicted by the models.  We
converted the Pan-STARRS1 apparent magnitudes to V-band using
taxonomy-dependent filter transformations if the asteroid's taxonomy
was specified by \citet{SDSS11} and, if not, the mean S+C class color
(see Table~\ref{tab.FilterTransformations}).

\begin{table}[!ht]
\begin{center}
\caption{Asteroid magnitude transformations from Pan-STARRS1 AB filter
  magnitudes to the Johnson-Cousin V- system based on
  \citet{Tonry12}. Solar colors are also included for reference.
  \label{tab.FilterTransformations}}
\vspace{20pt}
\begin{tabular}{c|cccccc}
Taxonomy & V-$g_{\rm P1}$ & V-$r_{\rm P1}$ & V-$i_{\rm P1}$ & V-$z_{\rm P1}$ & V-$y_{\rm P1}$ & V-$w_{\rm P1}$ \\
\hline
Sun & -0.217 & 0.183 & 0.293 & 0.311 & 0.311 & 0.114 \\
Q & -0.312 & 0.252 & 0.379 & 0.238 & 0.158 & 0.156   \\
S & -0.325 & 0.275 & 0.470 & 0.416 & 0.411 & 0.199  \\
C & -0.238 & 0.194 & 0.308 & 0.320 & 0.316 & 0.120   \\
D & -0.281 & 0.246 & 0.460 & 0.551 & 0.627 & 0.191    \\
X & -0.247 & 0.207 & 0.367 & 0.419 & 0.450 & 0.146   \\
\hline
Mean  (S+C) & -0.28 & 0.23 & 0.39 & 0.37 & 0.36 & 0.16  \\
\end{tabular}
\end{center}
\end{table}

The initial fits also use the mean class-dependent $G$ provided in Table~\ref{tab.G-vs-taxonomy} if the taxonomic class is specified in the SDSS
database \citep{SDSS11} but, if the class is not known, we use the
mean of the $S-$ and $C- $class values: $\overline{G}_B=0.15$ (B89)
and $\overline{G}_M=0.53$ \citep{Osk12} respectively.

\begin{table}[!ht]
\begin{center}
\begin{threeparttable}
\caption{Average slope parameters, $G_B$ and $G_M \equiv G_{12}$,
  adopted in this work for 5 asteroid taxonomic classes as measured by
  \citet{Pra12} and \citet{Osk12} respectively. The 6th row provides
  `standard' averages for the dominant S and C taxonomies.
  \label{tab.G-vs-taxonomy}
}
\begin{tabular}{c|c|c}
Taxonomic &  $G\equiv G_B$ & $G_{12}\equiv G_M$   \\
  Class   &  $\pm$(RMS)    & $\pm$(RMS)          \\
\hline
Q &  $0.25\pm0.13$       &  $0.41\pm0.14$      \\
S &  $0.24\pm0.06$       &  $0.41\pm0.16$      \\
C &  $0.15\pm0.09$       &  $0.64\pm0.16$         \\
D &  $0.09\pm0.09$       &  $0.47\pm0.14$        \\
X &  $0.20\pm0.09$       &  $0.48\pm0.19$        \\
\hline
S+C& $0.15$              &  $0.53$ \\

\end{tabular}
\end{threeparttable}
\end{center}
\label{tab.G-comparison}
\end{table}

The initial fits provided the absolute
magnitudes in both photometric systems, $H_{B,i}$ and $H_{M,i}$, that
were the inputs to the next step in the pipeline.

\subsection{Step 2:  Generating asteroid clones}
\label{sss.MonteCarloClones}

Our final $H$ and $G$ estimates are the result of Monte Carlo (MC)
simulations that require the generation of synthetic `clones' for each of the
asteroids in our sample.  Each of the clones is generated with its own
color, slope parameter, rotation period, light curve amplitude and
phase, where each of the parameters is selected from a unbiased distribution as described below.

\subsubsection{Clone colors}
\label{ssss.clone-colors}

Our pipeline  can assign each clone the color of its parent asteroid (if known) or,  when the parent's color is not known, a
random color based on an appropriate mix of taxonomies as a function
of semi-major axis.  About 16\% of the asteroids in our sample have taxonomies defined by \citet{SDSS11} (SDSS).

We implemented this technique by dividing the inner solar system into
4 zones (see table~\ref{Tab2}): NEO-like ($a<2\,au$), main belt ($2\,au
\le a < 3.7\,au$), Hildas ($3.7\,au \le a < 4.5\,au$) and Trojans
($4.5\,au \le a < 6.0\,au$).  The semi-major limits defining the zone edges were set at or near a
minimum in the number distribution as a function of semi-major axis and by the availability of published taxonomic
distributions.  The exact values make little difference to this
work. We used
the published, debiased taxonomic distributions in Table~\ref{Tab2} in the 4 zones with the qualification that for the main belt \citep{Diniz03} we aggregated many
related taxonomic types into 3 broad spectral classes: S-class=(A, AQ,
AV, O, OV, S, SA, SO, SQ, SV, V, L, LA, LQ, LS), X-class=(X, XD,
XL, XS), and Q-class=(Q, QO, QV). We required that the fraction, $f(c,z)$, of asteroids with
spectral class $c$ in zone $z$ satisfies $\sum_c f(c,z)=1$.  In the main belt, zone 2, we were able to generate the taxonomies as a finer function of $a$ as provided by \citep{Diniz03} with a similar requirement that $\sum_c f(c,a)=1$ at each semi-major axis.

\begin{table}[!ht]
\begin{center}
\begin{threeparttable}

\caption{Taxonomic distribution of asteroids in 4 semi-major axes
  zones used in this work.  The main belt values are given below at a
  representative $a=2.5\,au$ but we generated the clone
  taxonomies as a smooth function of semi-major axis in the range
  $2.0\,au \le a < 3.2\,au$ as specified by \citet{Diniz03}.}
\label{Tab2}
\begin{tabular}{c|cccc}
        & zone 1            &  zone 2          & zone 3                  & zone 4 \\
Taxonomy& NEO-like\tnote{a} &  MB\tnote{b}     & Hilda\tnote{c}          & Trojans\tnote{d}  \\
        & $a< 2\,au$         &  $a \sim 2.5\,au$ & $3.7\,au \le a < 4.5\,au$ & $4.5\,au \le a < 6\,au$\\
\hline
Q & 14 &  0   &  0 &  0 \\
S & 23 & 61   &  0 &  0\\
C & 10 & 30   &  7 & 10\\
D & 18 &  0   & 67 & 80 \\
X & 35 &  9   & 26 & 10\\
\hline
\end{tabular}
\begin{tablenotes}
\item [a] \citet{Stu04}
\item [b] \citet{Diniz03}
\item [c] \citet{Grav12}
\item [d] \citet{Grav12a}
\end{tablenotes}
\end{threeparttable}
\end{center}
\end{table}

\subsubsection{Clone slope parameters}
\label{ssss.clone-G}

We assigned slope parameters to the clones as a function of their
assigned taxonomic class ($c$).  i.e. the $k^{th}$ clone was assigned
a slope parameter $G_k(c) = \mathrm{ran[\overline{G}(c),\sigma_G(c)]}$
where $\mathrm{ran[x,y]}$ is a random number generated from a normal
distribution with mean $x$ and standard deviation $y$, and
$\overline{G}(c)$ and $\sigma_G(T)$ are the mean and RMS
of the distribution of slope parameters for class $c$, respectively
(Table~\ref{tab.G-vs-taxonomy}).

\subsubsection{Clone rotation periods, amplitudes and phases}
\label{ssss.clone-periods-amplitudes-phases}

The sparse Pan-STARRS1 data did not allow us to measure any asteroid's
rotation period and
light curve amplitude.  Furthermore, $<2$\% of the asteroids in our sample have measured light curves reported in the asteroid light curve database \citep[LCDB\footnote{\,The asteroid lightcurve database is publicly
    available at
    \url{http://www.minorplanet.info/lightcurvedatabase.html}};\ ][]{Warner09,Waszczak15}  The lack of this information introduces
systematic uncertainty and error into the absolute magnitude and slope
parameter determination.  We quantified these effects using our Monte
Carlo technique with synthetic sinusoidal light curves for each clone.

Asteroid brightness variations on the hours-to-days timescales are
usually caused by their non-spherical shape and rotation (the
exceptions are for the unusual cases where the phase angle changes
rapidly for close approaching NEOs, for multiple-systems in which
brightness changes can occur if the objects transit or eclipse each
other, and for objects with significant color variations).  We assumed that the observing geometry
(i.e. phase angle) effect on the asteroids' light curves are
negligible in the Pan-STARRS1 data because of the limited range in phase
angle coverage in our sample (Fig.~\ref{fig.PS1asteroids}).  For the
purpose of generating the clones' light curves we assumed that all the
objects are triaxial ellipsoids that generate simple sinusoidal light
curves with peak-to-peak amplitude $A$, period $P$, and rotation phase
$\theta$.  The offset from the unmodulated light curve at time $t$ is then $\Delta m(t) = A \sin( 2\pi t/P + \theta) / 2$.  

\begin{figure}[!ht]
\center
\includegraphics{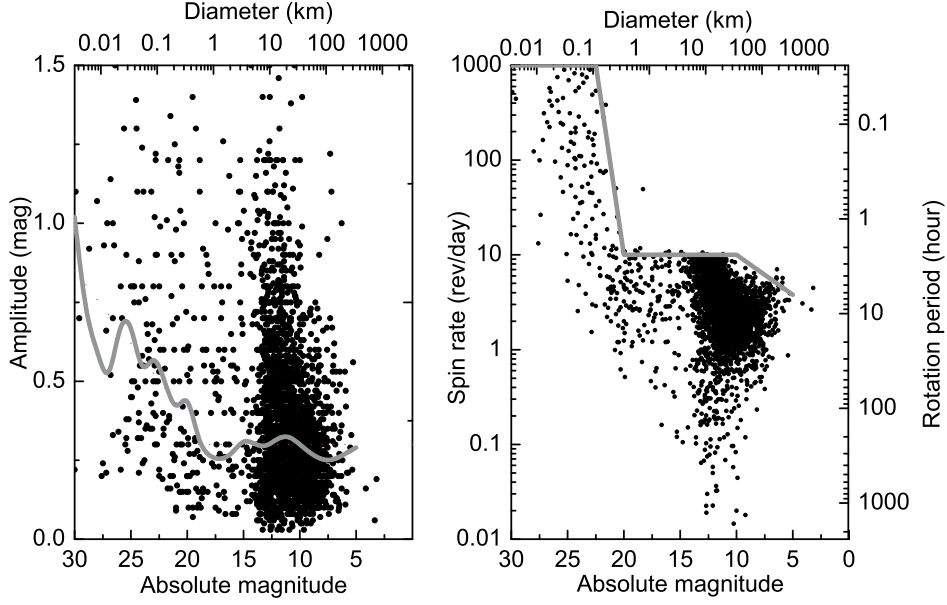}
\caption{(left) Asteroid light curve amplitudes vs. absolute magnitude
  ($H_B$) from the LCDB \citep{Warner09}. The solid gray curve
  represents the size-dependent moving median in 1.0\,mags wide bins.
  (right) Measured asteroid spin rates (periods are provided on the
  right) vs. absolute magnitude ($H_B$) from the LCDB
  \citep{Warner09}. The solid gray curve represents the size-dependent
  upper strength limit derived by \citet{Hol07}.}.
\label{fig.amp}
\end{figure}

Light curve amplitudes tend to be larger for smaller
asteroids \citep[see Fig.~\ref{fig.amp},\ ][]{Warner09}, probably
because the smaller objects tend to be more irregularly shaped.
Overall, the set of measured amplitudes and periods will be larger and shorter respectively than the true distribution because
of observational selection effects, larger amplitudes and shorter
periods are easier to detect and measure.

To reduce the impact of the light curve amplitude and period selection
effects we employed the debiased distributions derived by
\citet{Mas09} that are representative of asteroids with $H \sim 18$ (the average$\pm$RMS absolute magnitude in their study was $17.7\pm1.4$\,mags).  i.e. for objects with $H \sim 18$ they provide the cumulative fraction of asteroids, $F_{amp,Mas}(A)$, with light curve amplitudes $<A$.  We empirically estimate the cumulative distribution of light curve amplitudes at other absolute magnitudes $F_{amp}(A,H)$ by `normalizing' to the median at $H=18$ from the median at other values:
\begin{equation}
F_{amp}(A,H) = F_{amp,Mas} \Biggl[ A \times { A_{med}(18) \over A_{med}(H)} \Biggr]
\label{eq.F_amp}
\end{equation}
where $A_{med}(H)$ is an empirical function (Fig.~\ref{fig.amp}) representing the median amplitude of asteroids in the LCDB \citep{Warner09}.  Thus, given a clone's initial (\S\ref{sss.InitialFit}) absolute magnitude, $H_i$, we generated a random light curve amplitude for the clone according to the cumulative fractional distribution given by eq.~\ref{eq.F_amp}.

We followed a similar procedure in assigning each clone a rotation rate $R$ or, equivalently, a rotation period $P \equiv 1/R$.  \citet{Mas09} also provide the data from which we derive the cumulative fraction of asteroids, $F_{rot,Mas}(R)$, with rotation rates $<R$.  Once again, their results are representative of asteroids with $H \sim 18$, about 2\,mags fainter than the mean value in our data sample, so we developed an empirical technique to extend their cumulative fractional rotation rate distribution to other absolute magnitudes.

Asteroids with diameters $>100\,meter$ ($H < 23$) have an empirically observed upper limit to their rotation rate of about 12\,rev/day (Fig.~\ref{fig.amp}) and about 99\% of the distribution of debiased spin rates are $<12$\,rev/day \citep{Mas09}.  Asteroids larger than a few tens of kilometers ($H < 12$) have an even more restricted upper limit to their rotation rates.  We empirically defined an $R_{max}(H)$ as illustrated in fig.~\ref{fig.amp} and `compress' or `expand' the \citet{Mas09} distribution as necessary to create the cumulative fractional distribution at any $H$:  
\begin{equation}
F_{rot}(R,H) = F_{rot,Mas} \Biggl[ R \times { R_{max}(18) \over R_{max}(H)} \Biggr].
\label{eq.F_rot}
\end{equation}
Once again, given a clone's initial (\S\ref{sss.InitialFit}) absolute magnitude, $H_i$, we generated a random rotation rate for the clone according to the cumulative fractional distribution given by eq.~\ref{eq.F_rot}.

Finally, the rotational phase $\theta_k$ for the $k^{th}$ clone was generated
from a random uniform distribution in the range [0\,arcdeg,360\,arcdeg).

Our light curves were simple sinusoids even though we understand that real
asteroid light curves can be much more complicated.  The technique
could easily be extended to incorporate actual light curve properties
or a more realistic distribution but i) only a tiny fraction of known
asteroids have measured light curves ii) we will show below that
our results are not particularly sensitive to the actual light curve
parameter distribution and iii) if the actual light curve is known then there is no need for any of the methods developed here. i.e. this method only applies to the 98\% of asteroids that do not have measured light curves.  Since this is a preliminary work we have not made any effort to remove those asteroids that have published light curves.

\subsection{Step 3: Refining $H$ and $G$ (First Monte Carlo simulation).}
\label{sss.MC1}

The first Monte Carlo (MC) simulation yields our MC estimate for $H$
and $G$ from the sparse Pan-STARRS1 phase curve coverage data.  As
described in detail above, we created 500 clones of each object where
the $k^{th}$ clone was assigned a taxonomic class (color) $c_k$, light
curve amplitude $A_k$, and period $P_k$.  We then fit for each clone's absolute magnitude, slope parameter and
light curve phase, $(H'_k,G'_k,\theta'_k),$ by minimizing the $\chi^2$
with respect to the actual observations:
\begin{equation}
\chi^2_{k,obs} = \sum_{j=1}^{n} \;
 \Biggl[ 
   {
    m_k(t_j; H'_k, G'_k, \theta'_k) - m(t_j)
   \over
    \delta m(t_j)
   }
 \Biggr]^2
\label{eq.chi2_obs}
\end{equation}
where $n$ is the number of observations (detections) of the object, $m(t_j)$ is the actual object's observed apparent magnitude, $\delta m(t_j)$ is the reported uncertainty on the actual
Pan-STARRS1 apparent magnitude for that observation in the original
filter, and 
$m_k$ is the clone's predicted apparent magnitude at the actual time
of observation, $t_j$, in the Pan-STARRS1 filter in which the
  observation was made, with the clone's appropriate color
transformation ($\Delta m_k(t_j)$; Table~\ref{tab.FilterTransformations}):
\begin{equation}
m_k(t_j)=5\log[r(t_j)\Delta(t_j)]+\Phi[\alpha(t_j);H'_k,G'_k]+A_k\sin[2\pi t_j/P_k+\theta'_k]/2+\Delta m_k(t_j),
\label{eq.MC1mag}
\end{equation}
and $\Phi$ is the B89 or M10 phase function.

The `best' clone is the one ($k^*$) that
produces the minimum $\chi^2$ and we adopt that clone's $H'_{k^*}$ and
$G'_{k^*}$ values as our MC estimate for the object's absolute
magnitude and slope parameter.  The process was run separately for
both the B89 and M10 phase functions to provide our MC estimates for
$(H_B, G_B)$ and $(H_M, G_M)$ respectively. To avoid unphysical values
the fitting process required that $-0.25 \le G_B \le 0.8$ and $-0.5
\le G_M \le 1.5$.

We found that 500 clones provides a good balance between the
computation time and our ability to estimate the uncertainty on the
absolute magnitudes and slope parameters.  It is likely that when
there are only a small number of detections that the number of clones
could be decreased but we did not pursue this simplification.  When
the number of detections becomes very large then our technique becomes
unnecessary as either traditional \citep{Pra12} or sparse light curve
fitting \citep{Mui10,Law09} becomes more effective.

\subsection{Step 4: Estimating uncertainties and error on $H$ and $G$ 
(second Monte Carlo fit).}
\label{sss.MC2}

We estimated the uncertainties and errors on $H'_{k^*}$ and $G'_{k^*}$
by fitting for the absolute magnitude and slope parameter with purely
synthetic light curves generated from the clone with the best fit.
i.e. we re-applied the same method as described in Step 3
(\S\ref{sss.MC1}) except that we fit the clones to the best synthetic
object rather than the real object (we continue to use the sub-script
$k$ to refer to clones but the clones used here are distinct from the
clones used in the last step):
\begin{equation}
\chi^2_{k,syn} = \sum_{i=1}^{n} \;
 \Biggl[ 
   {
     m_k(t_j; H'_k, G'_k, \theta'_k) - m_{k^*}(t_j)
   \over
    \delta m_{k^*}(t_j)
   }
 \Biggr]^2.
\label{eq.chi2_syn}
\end{equation}
where $\delta m_{k^*}(t_j)=\delta m(t_j)$, i.e. the uncertainty on the synthetic observation at time $t_j$ was set to the uncertainty on the actual observation at time $t_j$.

If we let $X$ generically represent either $H$ or $G$ then the
combined statistical+systematic uncertainty on $X$ is the
standard deviation of the clones' $X$ distribution:
\begin{equation}
\delta X = \sqrt{ \frac{1}{n} \sum\limits_k {(X'_k - \overline{X'})}^2 }
\end{equation}
\noindent where $\overline{X'}$ is the average value of $X$ for all the
synthetic objects' clones.  Similarly, the combined
  statistical+systematic error on $X$ is the average error on the
values for the synthetic clones:
\begin{equation}
\Delta X = \frac{1}{n} \sum\limits_k { ( X'_k - X'_{k^*} )}
\end{equation}

\subsection{Verification}
\label{sss.Verification}

We verified our method with two independent sets of synthetic data generated from real Pan-STARRS1 data: 1) 10,000 randomly selected known Pan-STARRS1 objects, most of them with sparse phase curve coverage and 2) the 1,000 known Pan-STARRS1 objects with the best phase coverage.  To have better control over assessing our method's validity we generated photometric magnitudes and uncertainties with synthetic absolute magnitudes ($H_{B}$ and $H_{M}$) and slope parameters ($G_B$ and $G_M$) at each real time of observation with the known object's orbit. We then employed our pipeline to calculate each synthetic object's $H$ and $G$ to measure the statistical and systematic errors induced by our technique. Moreover, we tested two different scenarios for assigning light cure amplitudes and periods to the clones: 1) the debiased distributions from \citet{Mas09}, 2) and the observed distributions from the LCDB \citep{Warner09}. 

The result is that for both synthetic populations (sparse and dense phase curve coverage) and for both light curve amplitude-period relations (debiased and observed) the difference between the generated synthetic values and the values returned by our method was normally distributed with zero mean. i.e. our technique correctly derives the $H$ and $G$.  Use of the debiased or observed amplitude and period distributions does not affect the derived $H$ and $G$ at the level of photometric accuracy and uncertainty of the Pan-STARRS1 data with its associated phase curve coverage i.e. does not cause any systematic errors.

\section{Results \& Discussion}

\subsection{Absolute magnitudes comparison with \citet{Pra12}}
\label{sss.ComparisonWithPravec}

\begin{figure}[!ht]
  \begin{minipage}[b]{0.5\textwidth}
    \includegraphics[width=\textwidth]{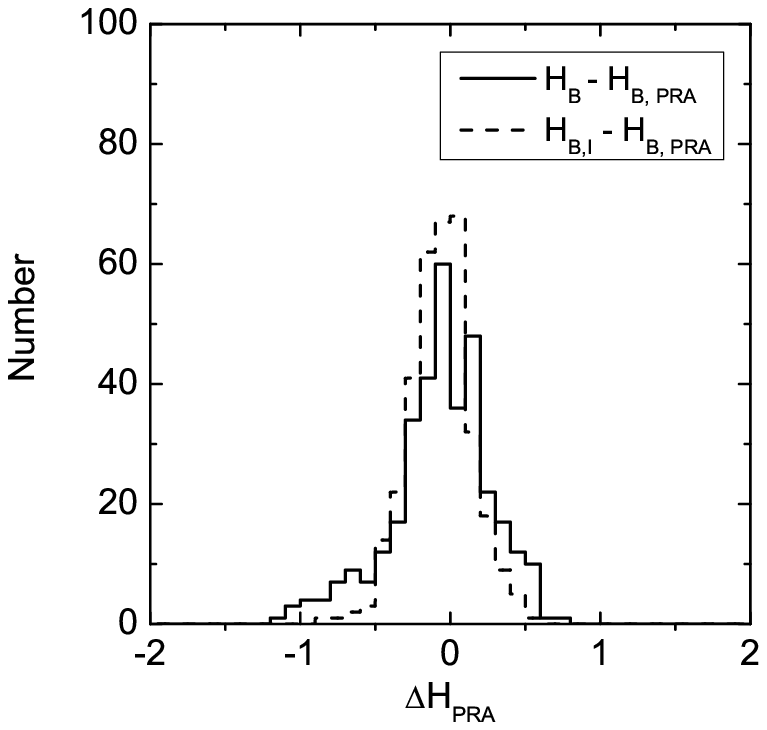}
 \end{minipage}
  \begin{minipage}[b]{0.5\textwidth}
    \includegraphics[width=\textwidth]{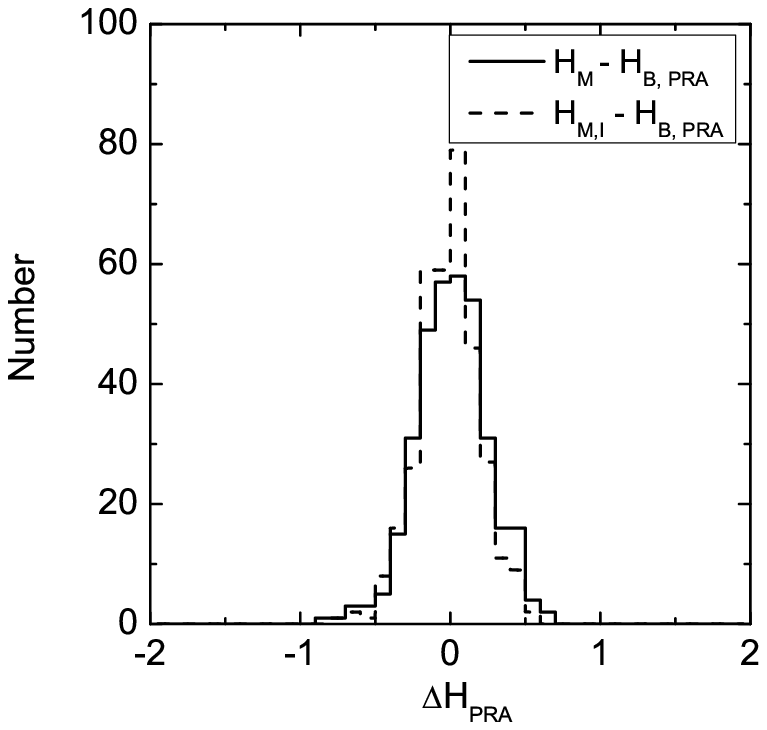}
  \end{minipage}
    \begin{minipage}[b]{0.5\textwidth}
    \includegraphics[width=\textwidth]{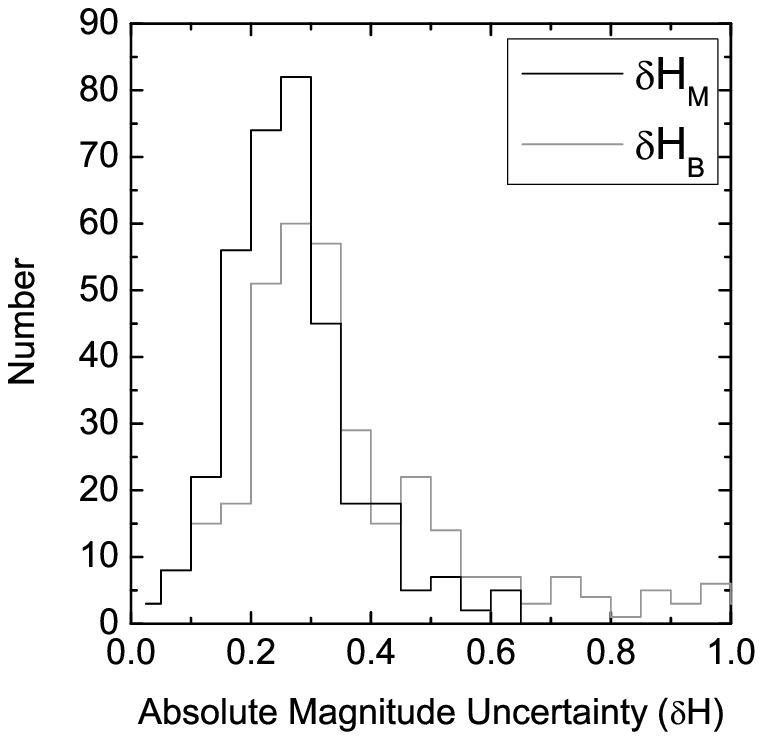}
  \end{minipage}
   \begin{minipage}[b]{0.5\textwidth}
    \includegraphics[width=\textwidth]{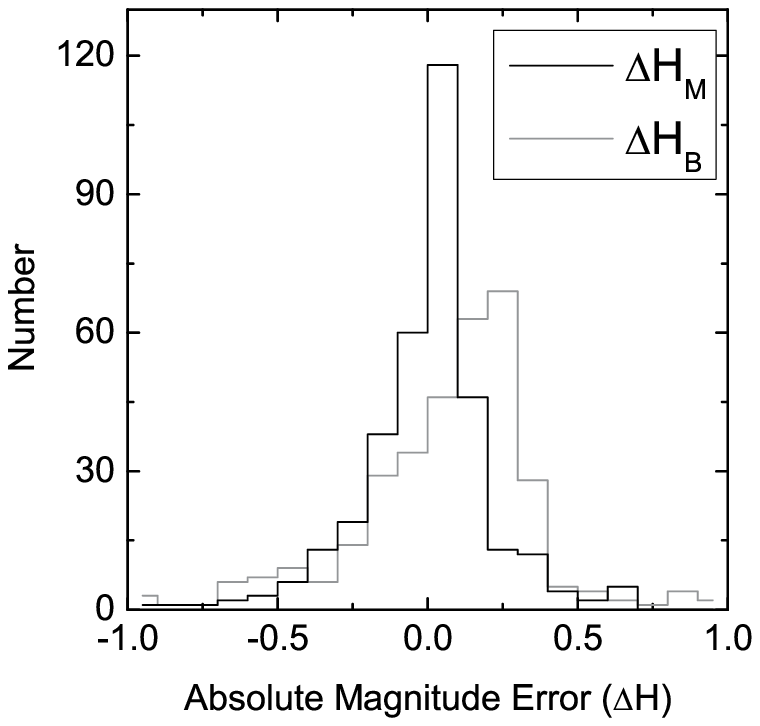}
  \end{minipage}

\caption{(top) Absolute magnitudes from our study compared with 347
  objects in common with \citet{Pra12} using the B89 (left) and M10
  (right) photometric systems.  The dashed line shows the results of
  the traditional initial fit (\S\ref{sss.InitialFit}) and the solid
  line provides the results of the MC fit
  (\S\ref{sss.MC1}). (bottom-left) Uncertainties and (bottom-right)
  estimated systematic errors on absolute magnitudes from our study
  compared with those reported by \citet{Pra12}.}
\label{fig.Pravec_comparisons}
\end{figure}

We think \citet{Pra12}'s detailed light curve study of $\sim500$
asteroids sets the standard in measuring 
asteroid photometric properties.  They provided only $H_B$ (it was
before the adoption of the new IAU standard) but that value {\it
  should} be identical to $H_M$.  Our results agree extremely
well with \citet{Pra12} for the 347 objects that appear in both data sets (fig.~\ref{fig.Pravec_comparisons}).  The mean differences of $H_B - H_{B,Pra} =
-0.06\pm0.02$\,mags and $H_M - H_{B,Pra} =
0.02\pm0.02$\,mags are consistent with zero to within $3\sigma$ and $1\sigma$ respectively, with better agreement for the new IAU standard photometric system of M10.  The RMS of each distribution is $0.36$\,mags and $0.29$\,mags respectively, due to the quadratic combination of the errors in both \citet{Pra12}'s and this work.

The distribution of $H_B - H_{B,Pra}$ is quasi-normally distributed
(fig.~\ref{fig.Pravec_comparisons}) with an RMS of $0.31$\,mags
including a tail extending to $H_B - H_{B,Pra}<-1$.  Interestingly,
the difference between our initial fits with assumed slope parameter
(\S\ref{sss.InitialFit}) and \citet{Pra12}, $H_{B,i} - H_{B,Pra}$, is
roughly normally distributed with a mean error of $-0.06\pm0.02$\,mags
and RMS of $0.26$\,mags.  Thus, the simple, traditional, fitting
procedure with assumed $G$ to our high-precision but sparse data
produces comparable absolute magnitudes to the MC technique.
The power of the MC technique lies in its ability to estimate the true
statistical and systematic uncertainty in the absolute magnitude due
to the unknown parameters in the analysis.

Our absolute magnitudes calculated with the M10 phase function ($H_M$)
are better behaved (fig.~\ref{fig.Pravec_comparisons}) in the
sense that the distribution is more normally distributed.  The initial
fit to the sparse data in the M10 system provided absolute magnitudes
with mean systematic errors of $0.00\pm0.02$\,mags and $\sigma\sim
0.26$\,mags compared to the MC technique with a mean error of
 $0.02\pm0.02$\,mags and $\sigma\sim
0.28$\,mags. The good behavior of both the MC and initial fits with M10
that results in a normal error distribution leads us to the conclusion
that it is superior for the determination of absolute magnitudes even
for sparse data samples.

We also used the \citet{Pra12} values to test our technique
(\S\ref{sss.MC2}) for establishing the uncertainty and error on our
measured absolute magnitudes. Their technique allows excellent control
of all the statistical and systematic uncertainties in the $H$
calculation because they observed targets for more than a decade in a
systematically controlled program and had 2 to 3 orders of magnitude
more data per object.  Thus, they report $H$ uncertainties about
$3\times$ less than our uncertainties and we can compare our measured
uncertainties ($\delta H$) to the RMS spread of $H - H_{Pra}$, and our
measured error estimates to its average
(fig.~\ref{fig.Pravec_comparisons}).

As stated earlier, the real power of the MC technique is
its ability to estimate the statistical and systematic uncertainties
on the derived $H$ and $G$ values. Our estimated absolute magnitude
uncertainties ($\delta H_B$; fig.~\ref{fig.Pravec_comparisons};
\S\ref{sss.MC2}) for the asteroids that overlap the \citet{Pra12} data
sample have the expected poissonian distribution with a mean of
$\bar\delta_{H_B}=0.36\pm0.01$\,mags using the B89 phase function
(fig.~\ref{fig.Pravec_comparisons}), comparable to the RMS
of $0.37\pm0.02$\,mags for the distribution of the error in our
measurement, $H_B - H_{B,Pra}$, as expected.  Similarly, our mean estimated systematic
error of $\Delta H_B = 0.03\pm0.02$\,mags agrees with the actual
systematic offset in the $H_B - H_{B,Pra}$ distribution.  We can
compare our estimated uncertainties and systematic errors in the same
manner for the M10 phase curve.  Our estimated mean uncertainty,
${\delta H_M}=0.26\pm0.01$\,mags, is consistent with $RMS(H_M -
H_{B,Pra})=0.28\pm0.02$\,mags and our estimated systematic error of
${\Delta H_M}=0.00\pm0.02$\,mags, is consistent with $\overline{(H_M -
  H_{B,Pra})}=0.02\pm0.02$\,mags.

The good agreement between our results and those of \citet{Pra12} illustrates the utility of our MC technique at measuring an asteroid's absolute magnitude and estimating the associated statistical+systematic uncertainty and any systematic bias, even for sparse data sets with limited phase angle coverage.  Furthermore, the nice behavior of our results with the M10 phase curve and the good agreement between our $H_M$ and $H_{B,Pra}$ provides evidence that $H_M \sim H_B$ when care is taken to ensure that the photometric data is of excellent quality.

\subsection{Absolute magnitudes}
\label{ss.AbsoluteMagnitudes}

\begin{figure}[!ht]
  \begin{minipage}[b]{0.33\textwidth}
    \includegraphics[width=\textwidth]{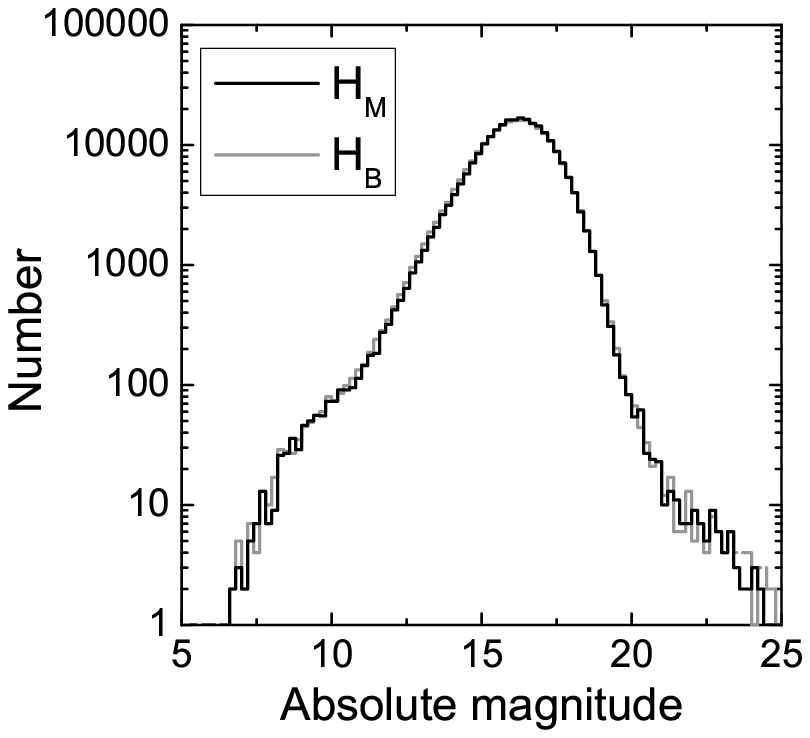}
 \end{minipage}
  \begin{minipage}[b]{0.33\textwidth}
    \includegraphics[width=\textwidth]{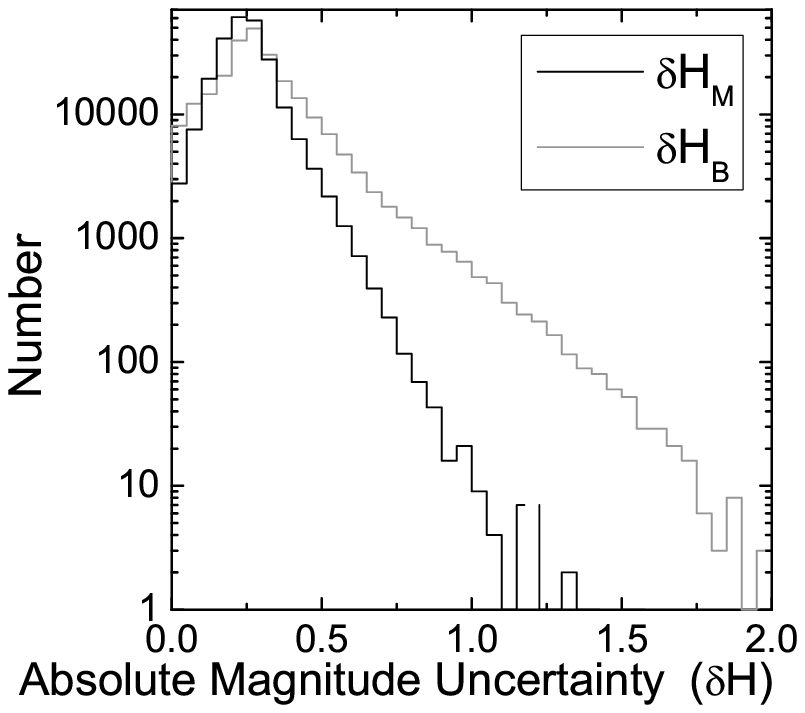}
  \end{minipage}
    \begin{minipage}[b]{0.33\textwidth}
    \includegraphics[width=\textwidth]{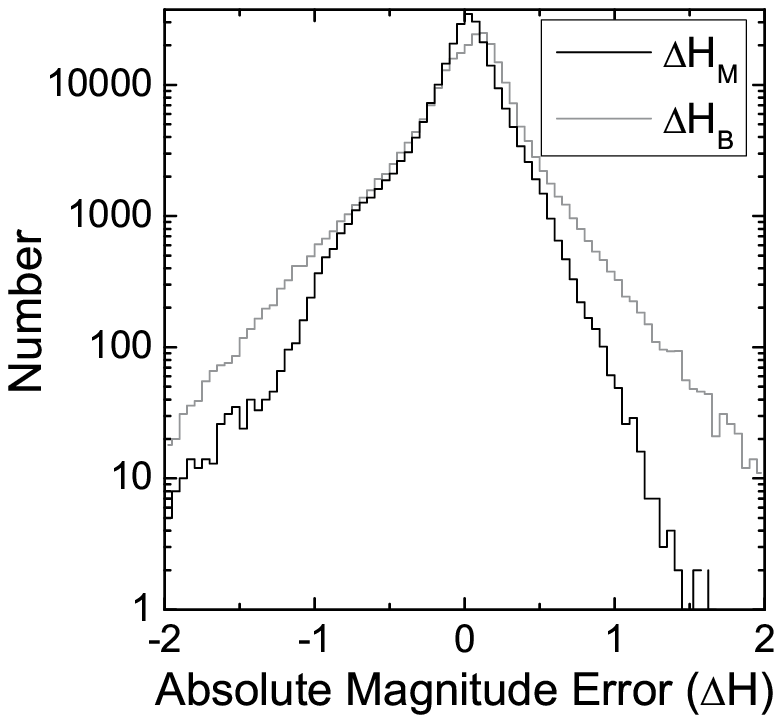}
  \end{minipage}

\caption{(left) Absolute magnitudes ($H_M$ and $H_B$) of 248,457
  asteroids.  (center) Uncertainties and (right) estimated errors in
  the absolute magnitudes derived with our Monte Carlo method using
  the phase functions of (gray) B89 and (solid) M10.}
\label{fig.H-distn-unc-err}
\end{figure}

Having established the utility of our technique on a well-controlled data set in the previous section we now employ it on all the asteroids in our selected Pan-STARRS1 data sample.  We were able to calculate the absolute magnitudes with combined
statistical and systematic uncertainties for more than 240,000 asteroids
spanning the range from $6.4 < H <
26.5$ (fig.~\ref{fig.H-distn-unc-err}).  The $\sim20$\,mags range corresponds to
about a factor of $10,000\times$ in the diameters of the objects and
spans the inner solar system from the NEOs to Jupiter's Trojan
asteroids.  Our sample represents $\sim38$\% of all known asteroids in
that range as of February 2014, with the highest completion of
$\sim75$\% from $10.5 < H < 11.0$.

\begin{figure}[!ht]
\begin{center}
\includegraphics{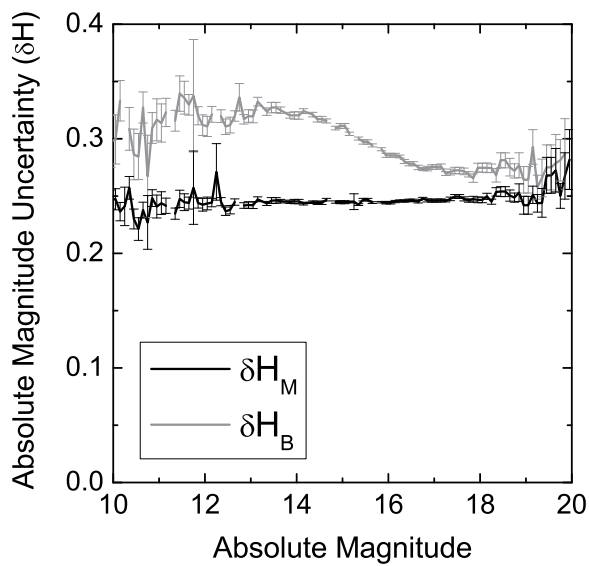}
\caption{Absolute magnitude uncertainty as a function of H using the B89 (grey) and M10 (dark) methods.}
\label{fig.HuncH}
\end{center}
\end{figure}

The mean uncertainties of $\overline{\delta H_B}=0.30\pm0.01$\,mags and
$\overline{\delta H_M}=0.25\pm0.01$\,mags (Fig
\ref{fig.H-distn-unc-err}) show that the new IAU photometric scheme of
M10 is better than B89 for the sparse Pan-STARRS1 data and
phase coverage but this conclusion mis-represents the full utility of the
M10 technique.  For one, the M10 system uncertainty is almost uniform
with $\delta H_M \sim 0.24$\,mags for the entire range $10<H<20$ (fig.~\ref{fig.HuncH}).
Second, even though the two techniques yield approximately the same
uncertainties for the faintest objects for which the uncertainty is
dominated by the measurement statistics (fig.~\ref{fig.HuncH}), the
B89 method's statistical uncertainty is $\sim 0.35$\% larger for bright objects ($10<H<14$).

The mean of our estimated statistical+systematic error using the M10 method, $\overline{|\Delta H_M|}=0.02\pm0.01$\,mags, is
comparable to the B89 method, $\overline{|\Delta H_B|}=0.01\pm0.01$\,mags (fig.~\ref{fig.H-distn-unc-err}).  The error
in the absolute magnitude for each asteroid is less than the estimated
uncertainty in $\sim62$\% of all the asteroids in our $H_B$ sample and $\sim73$\% in our $H_M$ sample.  The RMS of the $|\Delta H_B|$ and $|\Delta H_M|$ errors respectively of $\sim0.35$\,mags and $\sim0.25$\,mags confirms that the new IAU
photometric system is an improvement over the earlier one and,
furthermore, the shape of the error distribution is more reasonable
for $\Delta H_M$ than $\Delta H_B$ (note the peak of  $\Delta H_B$ is shifted by $0.05$\,mags from zero but the $\Delta H_M$  peak is near zero (fig.~\ref{fig.H-distn-unc-err}).

\begin{figure}[!ht]
  \begin{minipage}[b]{0.5\textwidth}
    \includegraphics[width=\textwidth]{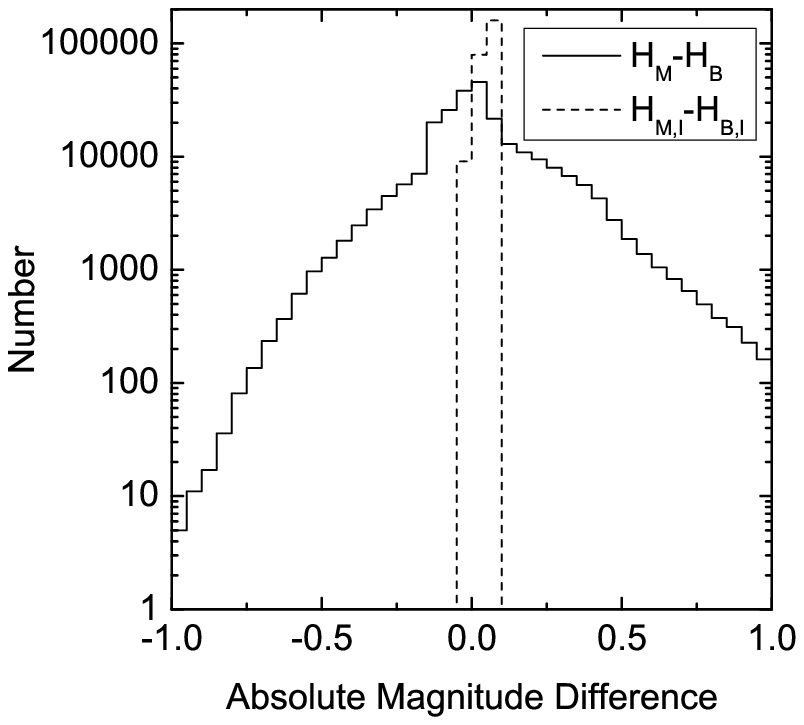}
 \end{minipage}
  \begin{minipage}[b]{0.5\textwidth}
    \includegraphics[width=\textwidth]{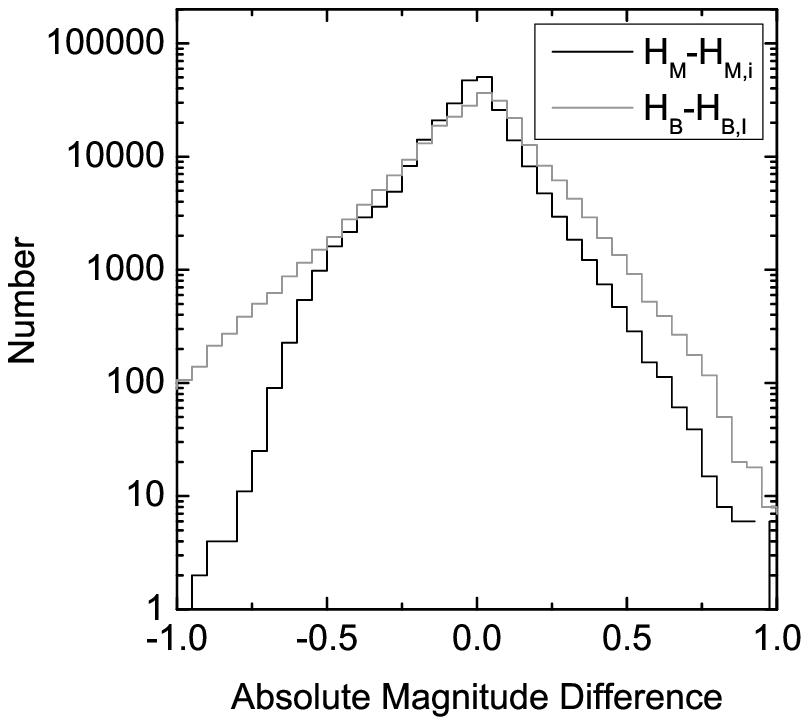}
  \end{minipage}
  \caption{
(left) Difference between the M10 and B89 absolute magnitudes for the MC and initial fit solutions. 
(right)  Difference between MC and initial fit solutions for the absolute magnitude using the M10 and B89 methods.}
\label{fig.deltaH}
\end{figure}

Overall, there is almost no difference between our M10 and B89 ensemble results for Pan-STARRS1 asteroids and the mean difference $\overline{H_M-H_B}$ is $0.03\pm0.01$\,mags with RMS of $0.22$\,mags (fig.~\ref{fig.deltaH}). The mean difference between the initial fit solutions is $\overline{H_{M,i}-H_{B,i}}=0.05\pm0.01$\,mags.

On the other hand, the utility of restricting $H$ and $G$ analyses to
data derived from well-calibrated single-survey data is easily
illustrated by comparing the results of our technique to the MPC
database values that do their best to incorporate data from multiple
telescopes and observers over many decades. The MPC currently only
publishes absolute magnitudes using the B89 phase function and there
is a mean difference of $\overline{H_M-H_{B,MPC}}=0.26\pm0.01$\,mags and
$\overline{H_B-H_{B,MPC}}=0.22\pm0.01$\,mags between our technique and
the MPC values.  The consistency between the mean differences is at
least reassuring and the RMS spread in values is due to 1) the
systematics introduced by the MPC's procedure that incorporates
apparent magnitudes from many different observatories in many
different passbands and 2) the systematics introduced by our sparse
light curve coverage.  Given that we established in
\S\ref{sss.ComparisonWithPravec} that our technique works well in
comparison to the `standard' \citet{Pra12} values, our conclusion is
that the error is due to the MPC's incorporation of photometry from
different sites and filters over a long period of time. The error
reported here is less than the $\sim 0.4$\,mags value reported by
\citet{Jur02}, but since the time of that study the MPC database has been
further populated by photometry from Pan-STARRS1 and other large surveys with
better photometric calibrations than previous surveys. Hence, it is
unsurprising that the $H_{B,MPC}$ values approach their correct
values over time.

\begin{figure}[!ht]
  \begin{minipage}[b]{0.5\textwidth}
    \includegraphics[width=\textwidth]{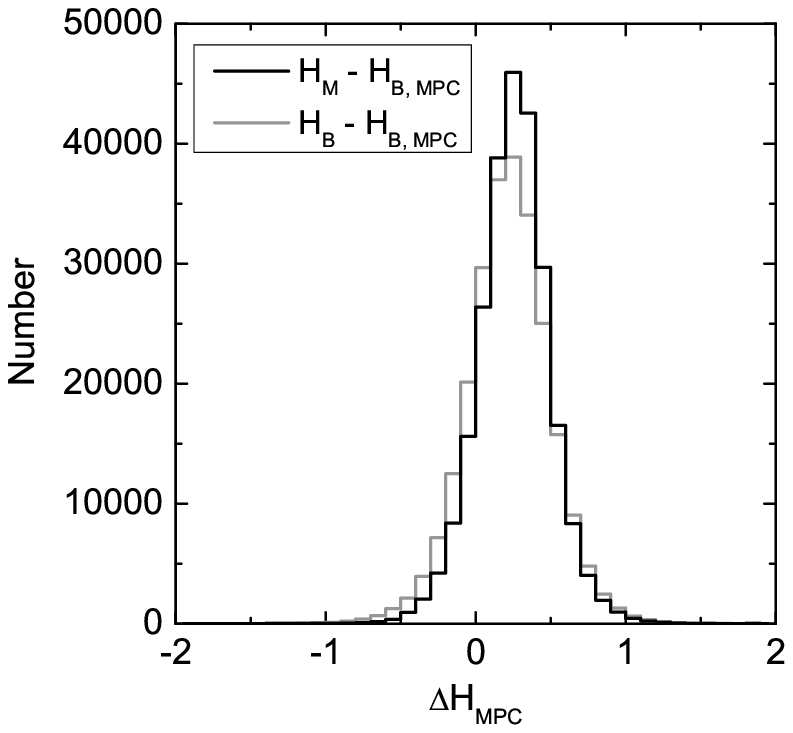}
 \end{minipage}
  \begin{minipage}[b]{0.5\textwidth}
    \includegraphics[width=\textwidth]{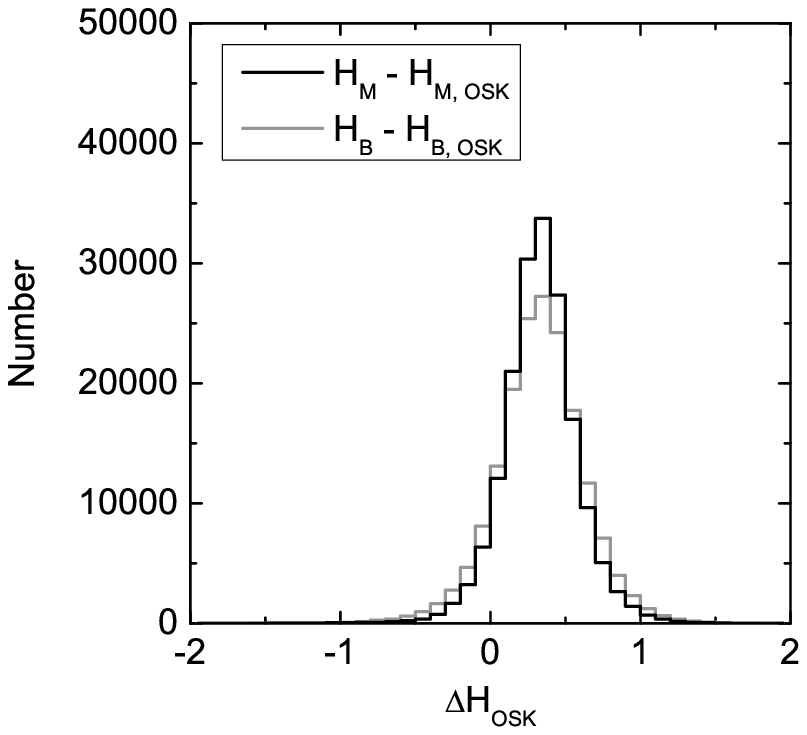}
  \end{minipage}
\caption{The difference between the $H_B$ and $H_M$ absolute magnitudes calculated in this work and (left) the Minor Planet Center and (right) \citet{Osk12}.  We compare our $H_M$ to the MPC $H_B$   because the two photometric systems should yield similar absolute magnitudes (in theory).}
\label{fig.HComparisons}
\end{figure}

Our calculated uncertainties are about $2\times$ larger than reported
by \citet{Osk12} who employed the entire MPC catalog for their input
photometry and provided $H_M$ and $G_M$ for 421,496 asteroids --- almost double our sample.  For comparison with earlier works they also provided $H_B$ and $G_B$.  Their work was very difficult as it required
calibrating and correcting the systematic problems intrinsic to the
various observatories and observers that contributed the photometric
data in multiple filters, but offered the advantage of an extensive data set with wide
time and phase angle coverage i.e. much like the MPC technique described in the last paragraph.  The systematic offset between our $H_M$ values and \citet{Osk12} of $H_M-H_{M,Osk}=0.33\pm0.01$\,mags (fig.~\ref{fig.HComparisons}) is similar to the offset derived between our results and the MPC.

\begin{figure}[!ht]
\center
\includegraphics{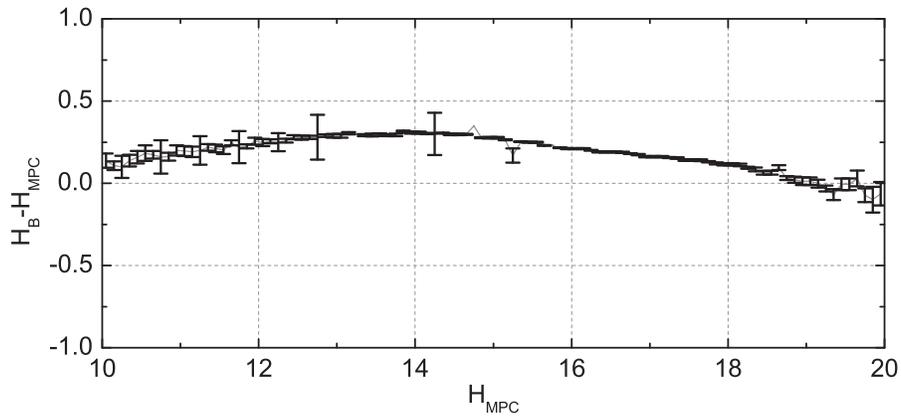}
\caption{The difference between the absolute magnitude $H_{MPC}$
  reported by the MPC using the B89 phase function and this work's
  $H_B$ value as a function of absolute magnitude ($\le 100$ random
  data points are shown at each discrete $H_{MPC}$ value in order to
  reduce confusion).  The thick solid black line represents the
  average in each $0.1$\,mags wide bin and the standard error on the
  mean is shown with error bars.  The error bars are about the width
  of the line for $13 <H_{MPC} <18$.}
\label{fig.MPC.vs.H.scatter}
\end{figure}

\citet{Jur02} and \citet{Pra12} reported a systematic offset of about
$0.38$\,mags to $0.5$\,mags between their calculated absolute magnitudes
and the values reported by the MPC. Those values are in rough
agreement with \citet{Waszczak15} who reported $H_B$ and $G_B$ from
over 54,000 asteroids observed in $g$ and $R$-band with the Palomar
Transient Factory (PTF). They measured a mean value of
$R_{PTF}=V_{MPC}+0.00$ which implies a systematic offset of
$\sim0.4$\,mags in the MPC absolute magnitudes because the average
$V-R$ for asteroids is $\sim0.4$.  Our values
(fig.~\ref{fig.MPC.vs.H.scatter}) are consistent with the MPC for
$H_B<11$\,mags and $H_B>19$\,mags, i.e. within $< 0.1$\,mags of the
MPC absolute magnitudes (their reported precision), but are
systematically higher than the MPC absolute magnitudes for 11\,mags$<
<H_B <19$\,mags.  i.e. our absolute magnitudes are
systematically {\it fainter} than reported by the MPC and this would
translate directly into predicting fainter apparent magnitudes than
the MPC and, similarly, suggesting that the objects are smaller than
predicted by the use of the MPC absolute magnitudes. The systematic
difference reaches a maximum of $\sim0.35$\,mags at $H_B\sim14$ in
agreement with the earlier studies.  This magnitude offset has
implications for developing observing programs, selecting objects for
followup, and for studies of the asteroids' size-frequency
distribution.

\clearpage
\subsection{Slope Parameters}
\label{ss.SlopeParameters}

The vast majority of Pan-STARRS1 asteroids offer only sparse phase angle
coverage (Fig.~\ref{fig.PS1asteroids}) for the determination of the
slope parameter but our MC technique should provide a realistic
estimate of the statistical uncertainty and systematic error when the
phase angle coverage is not too large and the detections are not in
multiple apparitions.  

\begin{figure}[!ht]
  \begin{minipage}[b]{0.5\textwidth}
    \includegraphics[width=\textwidth]{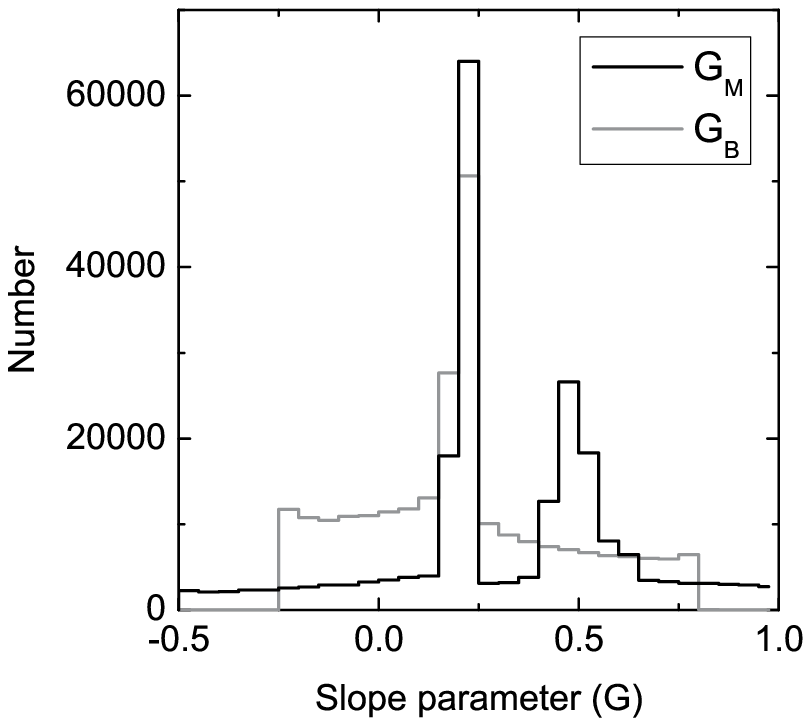}
 \end{minipage}
  \begin{minipage}[b]{0.5\textwidth}
    \includegraphics[width=\textwidth]{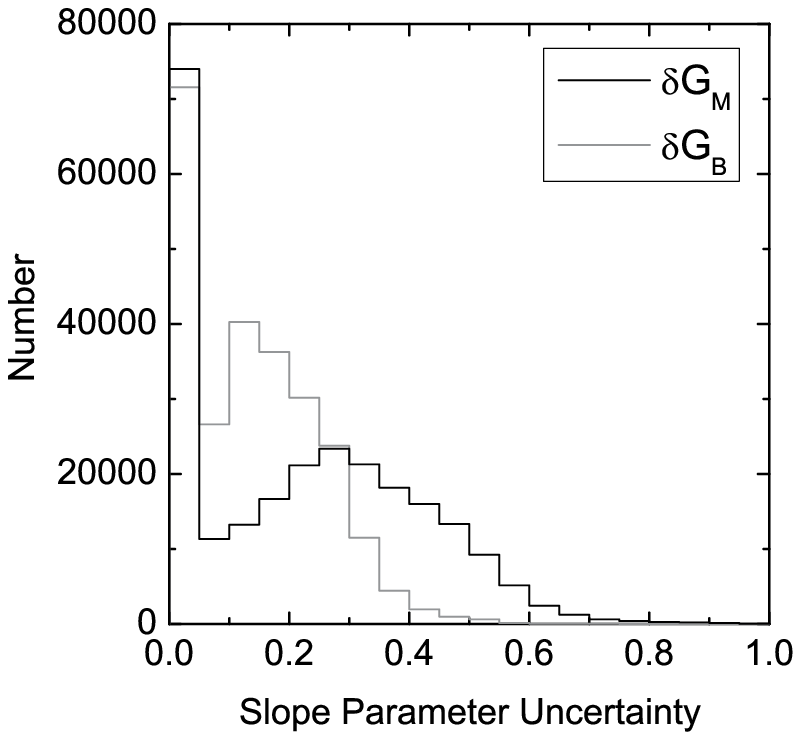}
  \end{minipage}
  \begin{minipage}[b]{0.5\textwidth}
    \includegraphics[width=\textwidth]{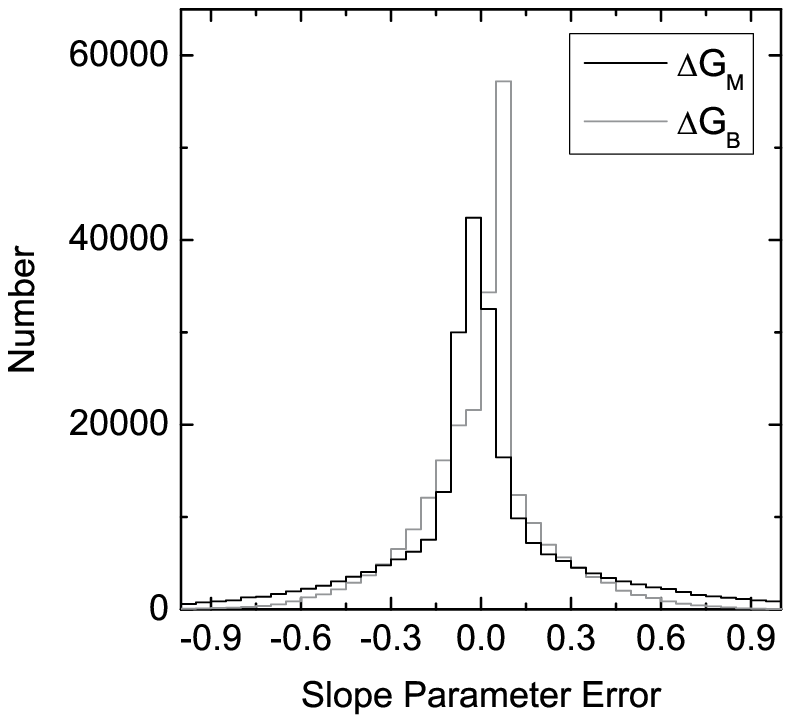}
 \end{minipage}
  \begin{minipage}[b]{0.5\textwidth}
    \includegraphics[width=\textwidth]{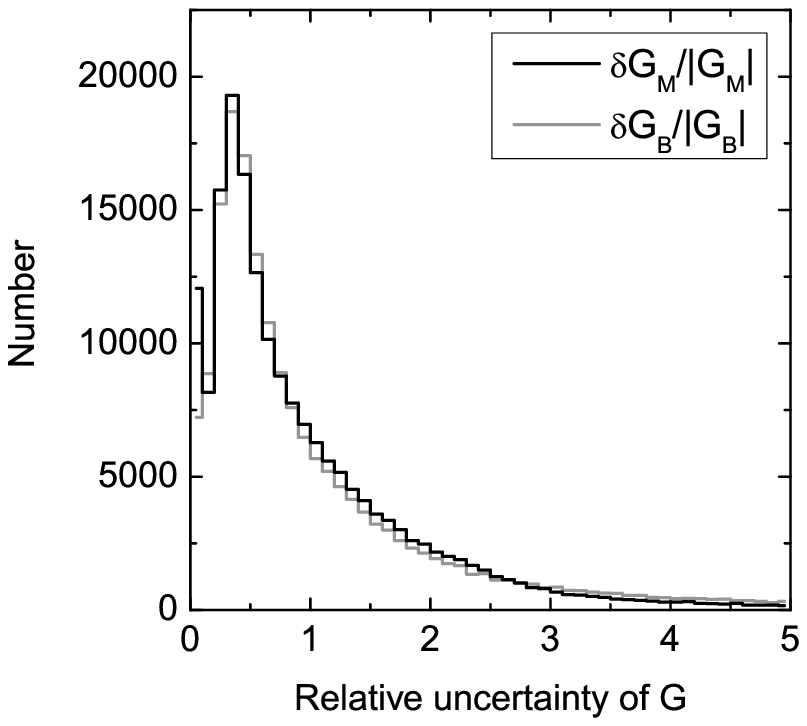}
  \end{minipage}
\caption{(top left) The slope parameters $G_B$ (B89) and $G_M$ (M10)
  and (top right) their uncertainties and (bottom left) errors.
  (bottom right) Percentage uncertainty in the slope parameter
  ($\delta G$) with the (gray) B89 and (black) M10 photometric
  methods.}
\label{fig.Gall}

\end{figure}

The $G_B$ distribution (fig.~\ref{fig.Gall}) is very wide with a peak
near 0.15, the default slope parameter for objects of unknown spectral
class (most of the asteroids in our sample). The distribution is
artificially constrained between the lower and upper limits
($-0.25<G_B<0.8$). On the other hand, the $G_M$ distribution has a
broad peak centered on $G_M\sim0.5$ superimposed on a roughly flat
distribution of slope parameters between our artificial limits
($-0.5<G_M<1.5$). The large peak near $G_M=0.2$ that contains
$\sim30$\% of all $G_M$ values is due to a discontinuity in the M10
phase function, it is not an error in our implementation.  In
comparison, $\sim8$\% of the \citet{Osk12} $G_M$ values were also
$\sim0.2$.  Our technique is particularly sensitive to the function
discontinuity and has a propensity to drive the fitted $G_M$ value to
0.2 when the number of data points is small. We suggest that future
attempts to use the M10 phase function flag and address this
situation, perhaps by forcing $G_M=0.5$ in those cases.

The slope parameter uncertainty ($\delta G$) distributions have peaks at zero corresponding to the $\sim24\%$ of cases in both methods where the MC technique did not converge and we fixed the slopes. Those $G_B$ that were actually fit have a normal-like distribution with mean $G_B = 0.18\pm0.01$ and RMS of 0.05 (Fig.~\ref{fig.Gall}).  Similarly, the $G_M$ uncertainty has a normal-like distribution with mean at $0.29\pm0.01$ and RMS of $0.17$. The $\delta G_M$  distribution is wider and shifted towards larger values than the $\delta G_B$ distribution because the $G_M$ values are fundamentally larger than the corresponding $G_B$ values. The percentage uncertainties ($\delta G/|G|$, fig.~\ref{fig.Gall}) in both slope parameters are very similar, suggesting that the two phase functions are equally effective for calculating the slope parameters, at least in the regime applicable to this data sample. The mean relative slope parameter uncertainties are $\sim 34$\% and $\sim 0.36$\% for $G_B$ and $G_M$ respectively, the large values being due mostly to the limited phase curve coverage.

\begin{figure}[!ht]
  \begin{minipage}[b]{0.5\textwidth}
    \includegraphics[width=\textwidth]{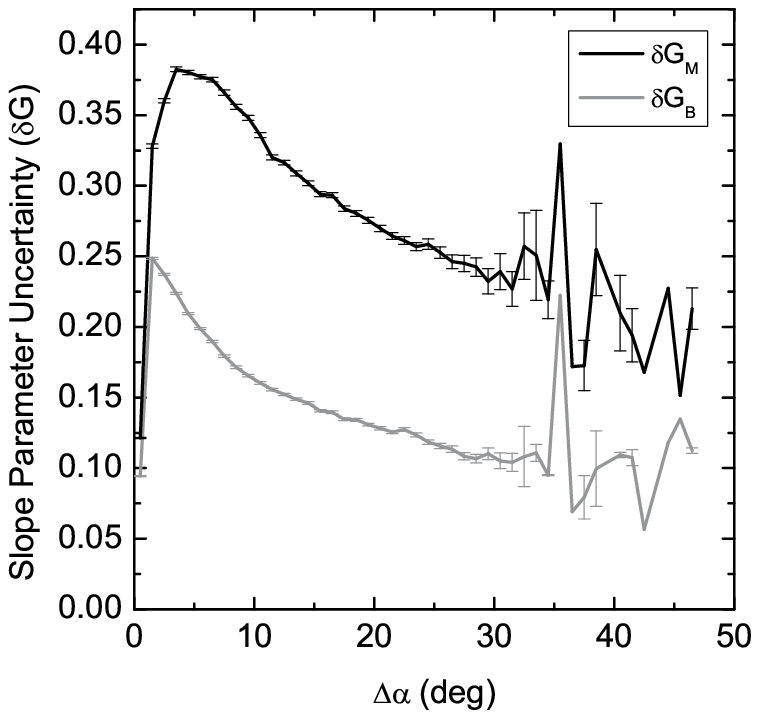}
 \end{minipage}
  \begin{minipage}[b]{0.5\textwidth}
    \includegraphics[width=\textwidth]{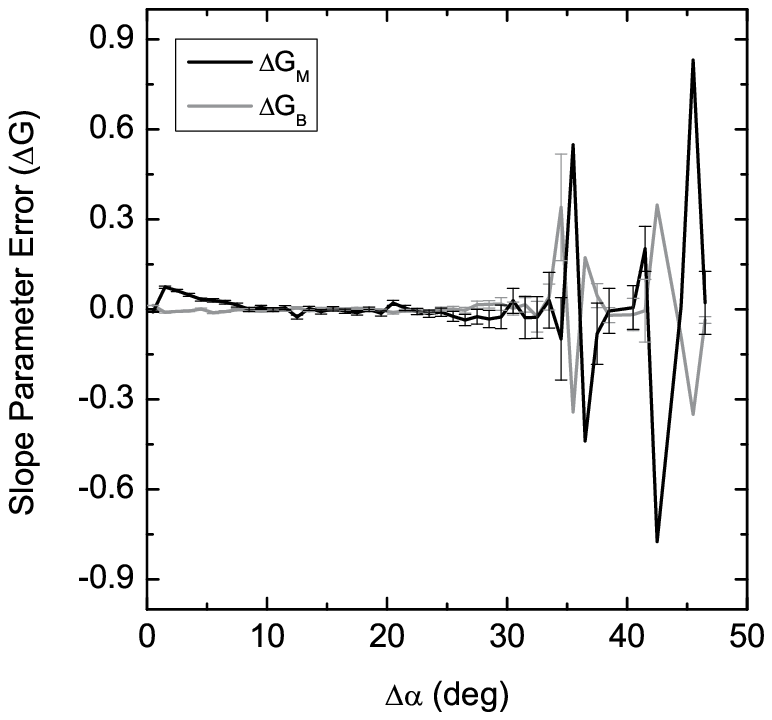}
  \end{minipage}
\caption{(left) Average MC slope parameter uncertainty and (right) error as a function of phase angle range using the (gray) B89 and (black) M10 phase functions.}
\label{fig.GUncDphase}
\end{figure}

As expected, the slope parameter uncertainty depends on the phase
angle coverage ($\Delta \alpha$, fig.~\ref{fig.GUncDphase}). The
uncertainty is artificially small at small phase angle ranges near
zero because in these cases the slope parameter was mostly fixed at a
pre-specified value. The uncertainty is largest for $\Delta \alpha
\sim 5$\,arcdeg because at this phase angle range the slope parameter
begins to be calculable, and the uncertainty drops at larger
phase-angle ranges because the data provides stronger constraints on
the shape of the phase function. However, even in the best case
scenario, for phase angle ranges of $> 30$\,arcdeg, the percentage
uncertainty is still $\sim50$\% for both phase functions. In any
event, the number of objects in our data sample with large phase angle
coverage is very small. Fig.~\ref{fig.GUncDphase} also illustrates
that the systematic errors introduced by our MC technique are not
dependent on phase angle coverage.

\begin{figure}[!ht]
  \begin{minipage}[b]{0.5\textwidth}
    \includegraphics[width=\textwidth]{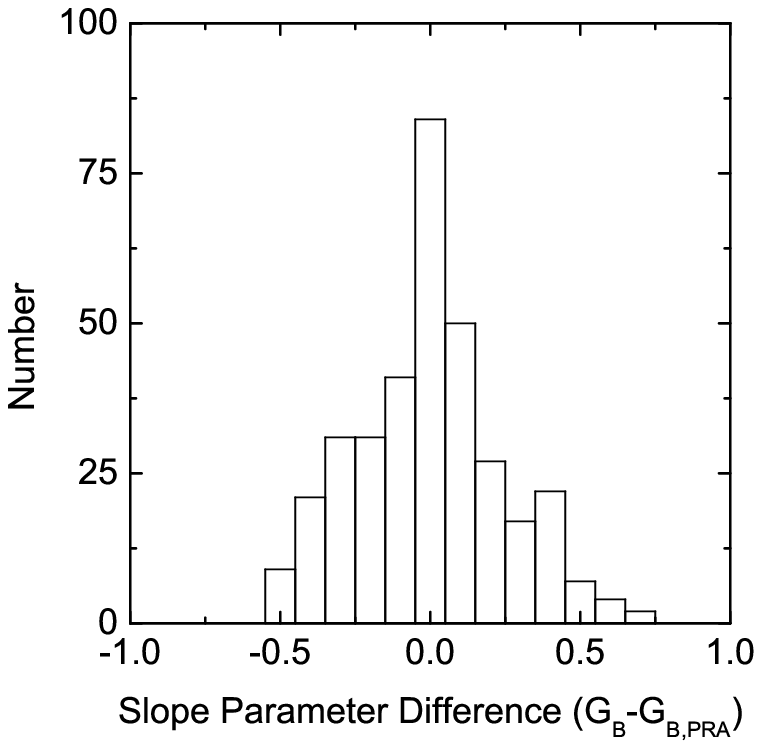}
 \end{minipage}
  \begin{minipage}[b]{0.5\textwidth}
    \includegraphics[width=\textwidth]{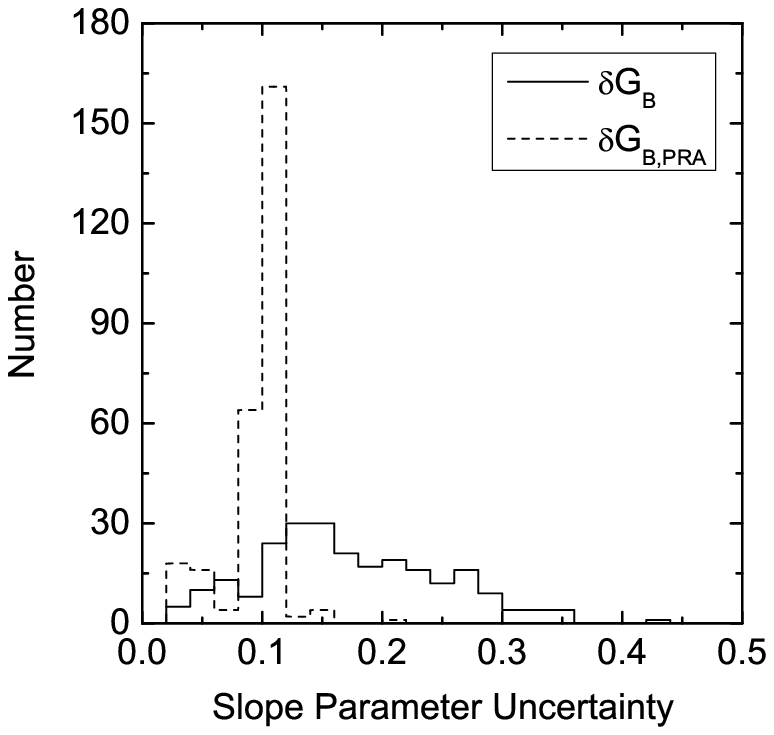}
  \end{minipage}
\caption{(left) The difference between our MC $G_B$ values (B89) and
  196 objects in common with \citet{Pra12}.  (right) Slope parameter
  uncertainty distributions for the same 196 objects for (solid) our
  MC values and (dashed) \citet{Pra12}.}
\label{fig.GPra}
\end{figure}

\citet{Pra12} provide acurate $G_B$ slope parameters with
uncertainties for more than 500 asteroids with densely covered light
curves in a single pass band over a wide range of phase angles. The
mean difference between this work's $G_B$ and $G_{B,PRA}$ is
$0.00\pm0.02$ with $\sigma\sim0.28$ for the 196 asteroids in common
between the two data sets with derived slope parameters
(Fig.~\ref{fig.GPra}). The agreement between our MC solution and the accurate work of \citet{Pra12} using the B89 phase function
suggests that our technique for calculating the slope parameter is
viable for a large number of asteroids with sparsely sampled light
curves. Furthermore, our technique allows us to estimate the mean
error on the derived slope parameter, $\overline{\Delta
  G_B}=0.00\pm0.01$, so the MC technique does not introduce a
systematic bias.  The mean statistical uncertainty in the slope
parameter for our data sample of $\overline{\delta G_B}=0.17\pm0.01$ is
twice as large as the \citet{Pra12} data set of $\overline{\delta
  G_B}=0.09\pm0.01$ which could be interpreted as either surprisingly
good, given the small number of observations and phase curve coverage
of our data sample, or as an indication that measuring $G_B$ is
difficult even with a very good data sample.

\begin{figure}[!ht]
  \begin{minipage}[b]{0.5\textwidth}
    \includegraphics[width=\textwidth]{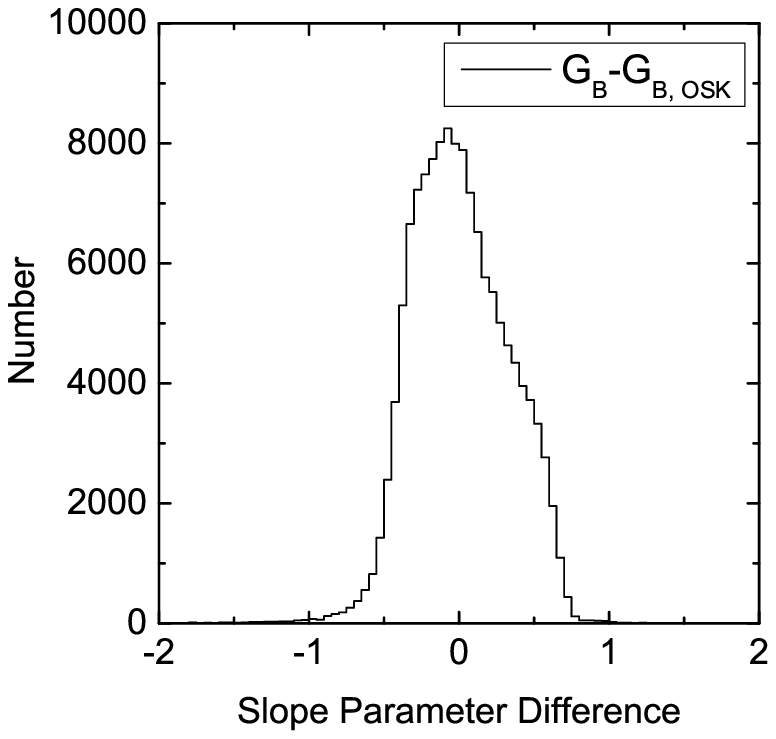}
 \end{minipage}
  \begin{minipage}[b]{0.5\textwidth}
    \includegraphics[width=\textwidth]{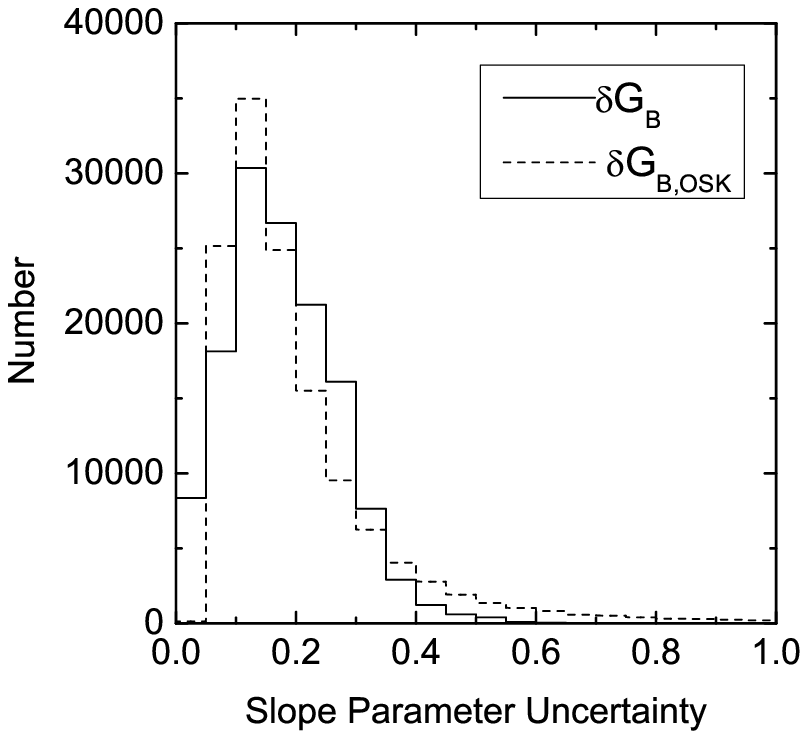}
  \end{minipage}
  \begin{minipage}[b]{0.5\textwidth}
    \includegraphics[width=\textwidth]{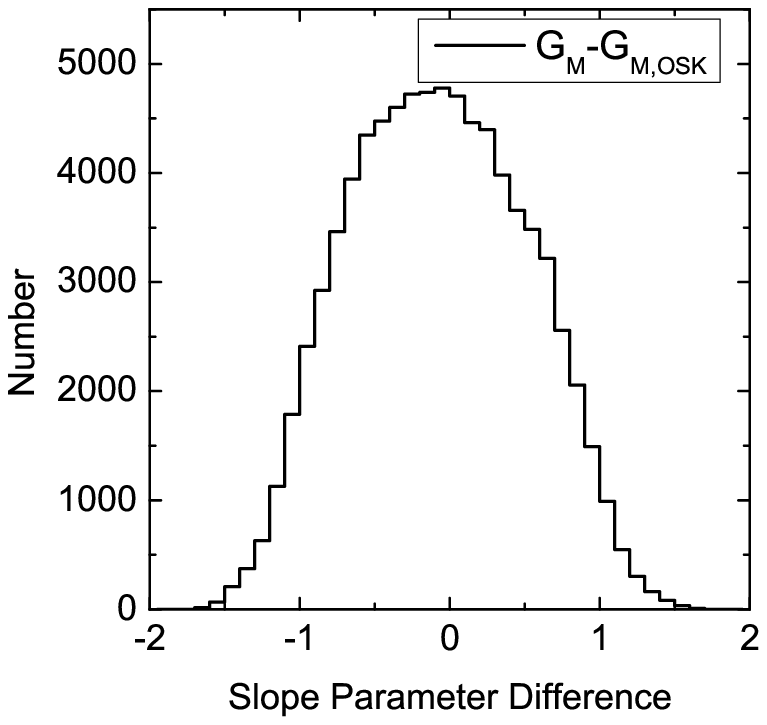}
 \end{minipage}
  \begin{minipage}[b]{0.5\textwidth}
    \includegraphics[width=\textwidth]{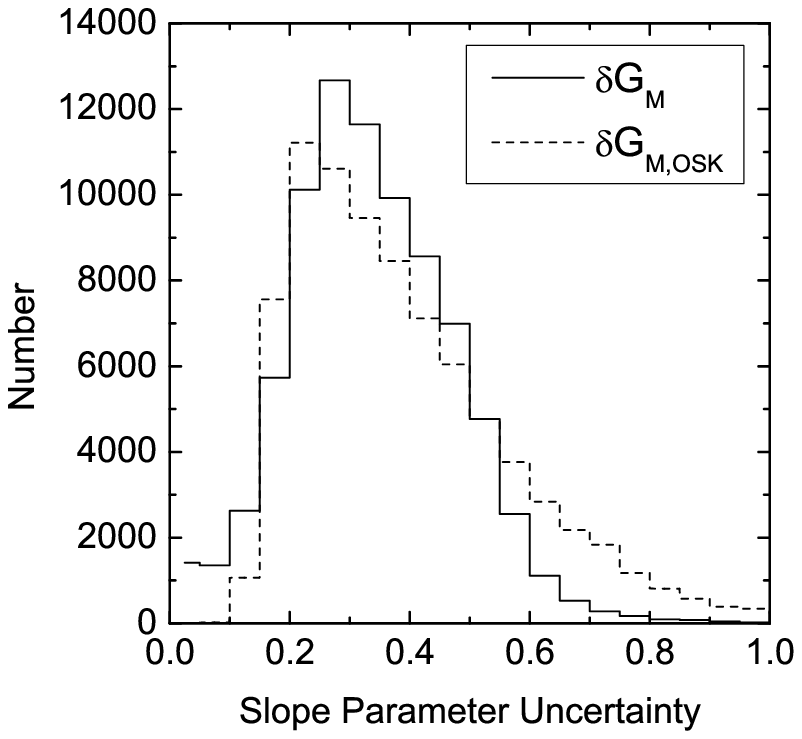}
  \end{minipage}
\caption{(top left) Difference between our MC $G_B$ values (M10) and 133,885 objects in common with \citet{Osk12}. (top right) Slope parameter uncertainties for the same objects as determined in this work and by \citet{Osk12}. (bottom left) Difference between our MC $G_M$ values (M10) and 80,756 objects in common with \citet{Osk12}. (bottom right) Slope parameter uncertainties for the same objects as determined in this work and by \citet{Osk12}.}
\label{fig.GOsk}
\end{figure}

\begin{table}[!ht]
\begin{center}
\begin{threeparttable}
\caption{Mean slope parameters $\pm$ standard deviation ($G_B$, B89) derived
  in this work (PS1, second column) and by \citet{Pra12} (PRA12, third
  column) for the same objects in 4 major taxonomic classes. The last
  column is the number of common objects that have a \citet{SDSS11} spectral classification (no D type asteroids satisfied our requirements on
  taxonomic identification and slope parameter determination).}
\begin{tabular}{c|>{\bf}cc|c}
Taxonomic  &  $G_B$  & $G_B$         &     \\
   Class   &  PS1    & PRA12         &  N  \\
\hline
Q  & 0.11$\pm$0.16 & $0.19\pm0.10$   &   3 \\
S  & 0.16$\pm$0.26 & $0.23\pm0.05$   &  32 \\
C  & 0.03$\pm$0.10 & $0.13\pm0.01$   &   4 \\
D  &     n/a       &     n/a         &   0 \\
X  & 0.21$\pm$0.30 & $0.20\pm0.10$   &   9 \\
\hline
\end{tabular}
\end{threeparttable}
\end{center}
\label{tab.GPRAtaxonomy}
\end{table}

\begin{table}[!ht]
\begin{center}
\begin{threeparttable}
\caption{Slope parameters derived in this work (PS1: $G_B$, second
  column; $G_M$ fifth column) and by \citet{Osk12} (OSK12: $G_B$,
  third column; $G_M$, sixth column) for the same objects in five
  different spectral classes. The forth and last columns are the number
  of objects in common between the two data sets with SDSS
  spectral classification \citep{Carvano10}.}
\label{tab.GOSKtaxonomy}
\begin{tabular}{c|>{\bf}ccc|>{\bf}ccc|c}
Taxonomic   &  $G_B$   & $G_B$      &N   & $G_M$     & $G_M$          &   N\\
  Class     &  PS1  & OSK12& & PS1   & OSK12 &  \\
\hline
Q & 0.21$\pm$0.28 & $0.20\pm0.10$ & 1324& 0.46$\pm$0.53 & $0.54\pm0.22$ &   886 \\
S & 0.22$\pm$0.28 & $0.19\pm0.22$ & 14686&0.47$\pm$0.53 & $0.55\pm0.20$ & 10231 \\
C & 0.18$\pm$0.28 & $0.16\pm0.10$ & 7892& 0.58$\pm$0.55 & $0.66\pm0.23$ &  5150 \\
D & 0.23$\pm$0.29 & $0.19\pm0.12$ & 1321& 0.42$\pm$0.52 & $0.61\pm0.25$ &   852 \\
X & 0.19$\pm$0.28 & $0.18\pm0.11$ & 2073& 0.53$\pm$0.54 & $0.59\pm0.24$ &  1428 \\
\hline
\end{tabular}
\end{threeparttable}
\end{center}
\end{table}

As described earlier, \citet{Osk12} derived asteroid slope parameters
from photometry reported to the MPC from multiple observatories that
used different filters and reference catalogs. They also had to deal
with the fact that the MPC observation submission format did not allow
reporting of photometric uncertainties. To reduce some of the
associated problems they statistically calibrated the disparate
datasets and used photometry only from major surveys. After excluding
the artificial peak near $G_M=0.2$ (i.e., excluding the range
$0.18<G_M<0.22$), and including only those objects for which $G$ was
actually fit, there were 80,756 objects in common with our $G_M$
values and 133,884 objects for comparison with our $G_B$.  The
wide and oddly-shaped distribution of the difference in slope parameters between our MC
technique and \citet{Osk12} (fig.~\ref{fig.GOsk}) illustrates the difficulty and large uncertainty in measuring $G$. The 
distribution peaks at zero for the B89 phase function with 
($\overline{G_B-G_{B,OSK}}=0.00\pm 0.01$) but there is a significant
offset using the M10 phase function of
$\overline{G_M-G_{M,OSK}}=-0.06\pm0.01$
(fig.~\ref{fig.GOsk}).  The RMS of the difference is larger using the
M10 ($0.58$) than with the B89 phase function ($0.35$) but this is
expected due to the numerically larger expected values of
$G_M\sim0.5$.

Fig.~\ref{fig.GOsk} also illustrates that our MC technique yields
slope parameters that are comparable or marginally better than the
work of \citet{Osk12}, even though our data sample includes much less
photometric data per object over a narrower phase angle range,
presumably because of the Pan-STARRS1 system's superior photometry and the
use of measured photometric uncertainties.  The
mean uncertainty for 80,756 objects in common with \citet{Osk12} is
$0.33\pm0.01$ (RMS$=0.14$) with our MC technique and is $0.39\pm0.01$
(RMS$=0.18$) for the values reported by \citet{Osk12}.

Slope parameters are taxonomy-dependent
\citep{Har89,Lag90,Osk12,Pra12} but most of the objects in our
Pan-STARRS1 data sample are fainter than known asteroids with well
established taxonomies, so we relied on the SDSS spectral
classification \citep{SDSS11} to assess our method's ability to detect
the taxonomic-dependence. We found 48 asteroids in common with
\citet{Pra12} and 18,541 with \citet{Osk12} (excluding values around
$G_M\sim0.20$) for which we could compare our calculated slope
parameters. Our mean$\pm$RMS $G_B$ values are consistent with
\citet{Pra12} (Table~\ref{tab.GPRAtaxonomy}) but our RMS distribution
is much larger and the common number of asteroids is very
low. Similarly, our $G_B$ and $G_M$ values
(Table~\ref{tab.GOSKtaxonomy}) are consistent \citet{Osk12} but the
RMS is distributions are large in both cases. There is a formal
difference between the means of some of the taxonomic classes but we
do not consider them further because of the large uncertainties on
each value and the large RMS of each taxonomic class' $G$
distribution.

As discussed above in section \ref{sss.InitialFit}, the phase curve coefficients $G_B$ and $G_M$ are
functions of asteroid composition. Given the compositional trends of the inner
belt being dominated by silicate S-type asteroids and carbon/volatile-rich
asteroids in the outer belt, we should  expect to see these trends reflected in our
derived phase functions. A similar study was performed by \cite{Osk12} in their
analysis of the MPC database. They found correlations between their measured G12 and orbital
elements throughout the main belt, reflecting the general compositional gradient and
family structure. To study this in our database we selected the 51,864
asteroids with orbital semi-major axes $2.1\leq a \leq 3.3$ AU where the range of phase angles
observed was $\Delta \alpha \geq 5^\circ$ and there were $N \geq 6$ observations. We then
calculated the running median values $\overline{G_B}$ and $\overline{G_M}$ as a function
of orbital $a$ over a range $\Delta a=0.05$ AU.

\begin{figure}[!ht]
  \begin{minipage}[b]{0.5\textwidth}
    \includegraphics[width=\textwidth]{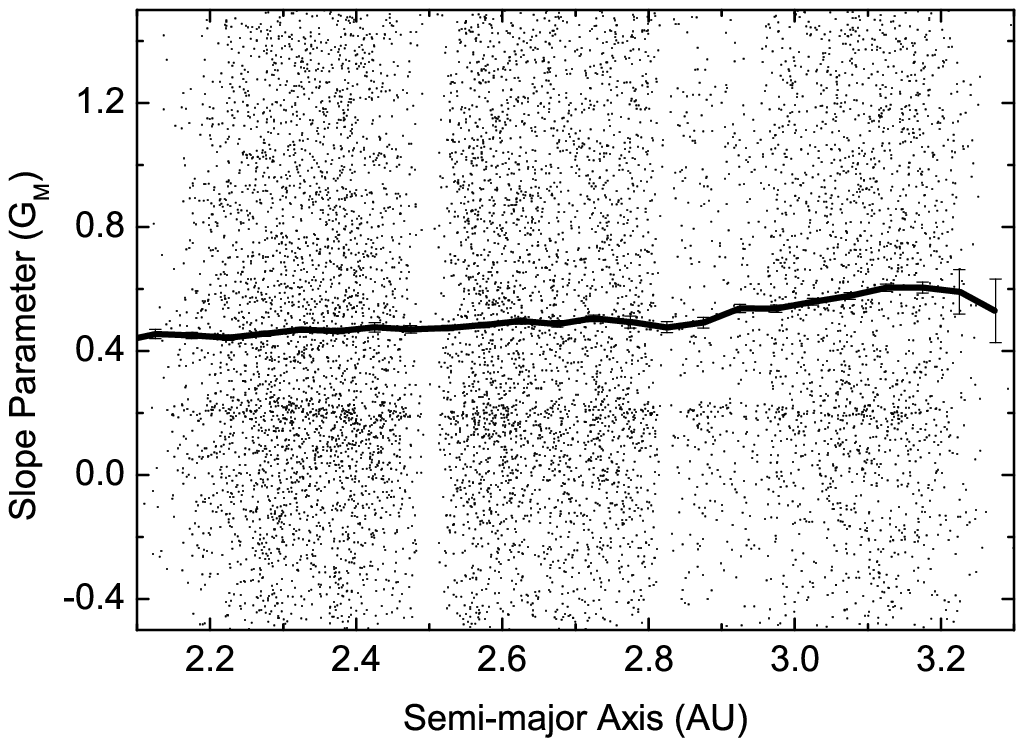}
 \end{minipage}
  \begin{minipage}[b]{0.5\textwidth}
    \includegraphics[width=\textwidth]{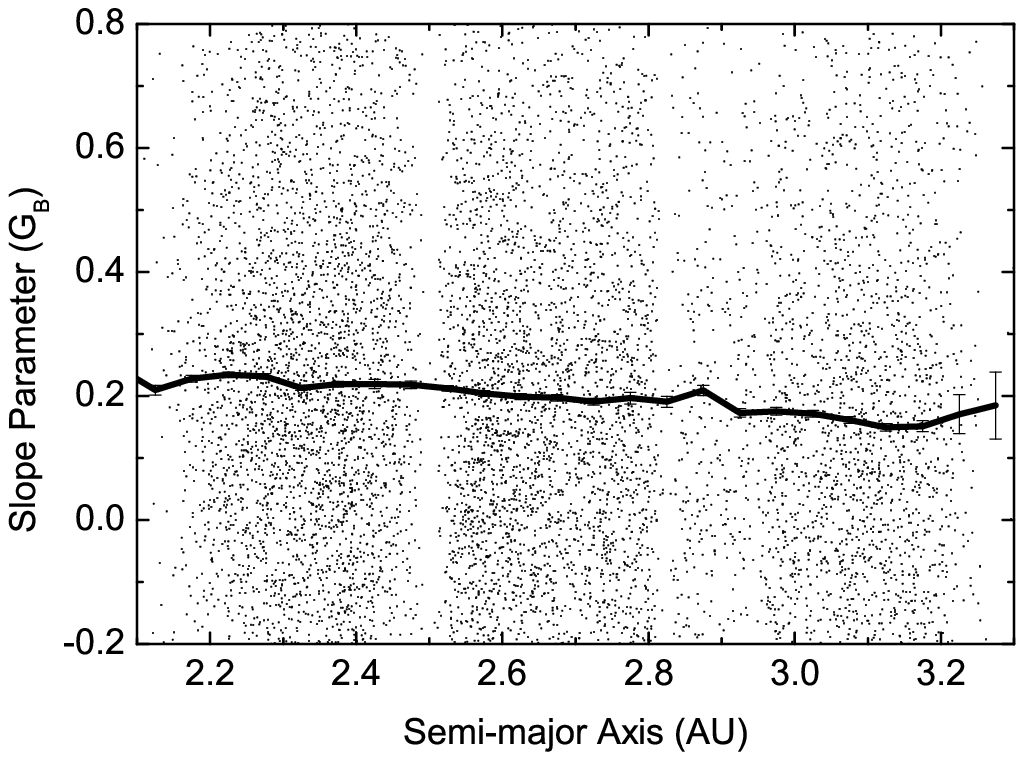}
  \end{minipage}
\caption{Moving average of $G_M$ (top) and $G_B$ (bottom) as a function semi-major axis.}
\label{fig.Gtrend}
\end{figure}

Figure \ref{fig.Gtrend} clearly shows clear a negative trend in $\overline{G_B}$
and a positive trend in $\overline{G_M}$ with orbital $a$. As $G_B$ is larger for S-type
than C-type asteroids, while $G_M$ becomes smaller, this agrees with the
established compositional gradient in the main-belt. For modelling purposes,
these trends may be approximated by the relationships
$\overline{G_B}=-0.103a + 0.446$ and $\overline{G_M}=0.237a - 0.175$ within the main belt. 
The largest deviations from these relationships occur at the 3:1 Kirkwood gap at 2.50 AU,
and at the 7:3 gap at 2.95 AU. This latter position marks where the S-type asteroids of
the dominant Koronis family of gives way to the T/X/K/D-type asteroids of the Eos family
\citep{Mothe05}.
We note that the overall observed scatter in individual values is dominated by $\Delta G$,
although it will also be partly due to the large amount of compositional mixing present
in the main belt \citep{DeMeo13}. 

\
\section{Availability}

The Pan-STARRS1 absolute magnitudes and slope
parameters with associated uncertainties as described herein are
available on-line (Appendix~\ref{appendix}).  The eventual goal is that the 
catalog will be updated with all the data from the entire 3 year
Pan-STARRS1 mission and then updated regularly with new data from the
ongoing extended mission that is purely focused on the solar
system. This effort will provide almost complete coverage of all known
asteroids with extensive phase angle coverage and good number of
detections per object.

\section{Conclusions}

Our work introduces a Monte Carlo method for
calculating absolute magnitudes ($H$) and slope parameters ($G$) {\it and their statistical uncertainties and systematic errors} that is applicable to single apparition asteroid observations and designed to handle limited photometric data over a restricted phase angle range.  The technique's utility was confirmed by comparing our $H$ and $G$ values to the well-established results of \citet{Pra12} for a limited number of objects.
We then applied it to derive $H$ and $G$ with statistical
uncertainties and systematic errors for $\sim240,000$ numbered
asteroids observed in the first 15 months of Pan-STARRS1's 
3-year nominal mission. The single-survey data, consistent
image processing, and well-defined photometric calibration, eliminates many of the problems encountered in past attempts to measure absolute magnitudes and slope parameters from a combination of different surveys. 

We find that the \citet{Mui10} phase function provides better results than the \citet{Bow89} phase function in terms of reducing the statistical uncertainty and systematic error on the absolute magnitude --- both crucial to accurately predicting ephemeris apparent magnitudes and calculating asteroid albedos from $H$ and measured asteroid diameters. There is a systematic $H$-dependent offset between the Minor Planet Center's reported absolute magnitude and $H$ derived in this work with a maximum offset of about $0.25$\,mags at $H\sim14$. 

The measured slope parameters are generally in agreement with the results of \citet{Pra12} and \citet{Osk12} but the statistical uncertainty and systematic error on any individual asteroid's $G$ is large due to poor temporal and phase-space coverage.

\section{Acknowledgements}

The Pan-STARRS1 Surveys have been made possible through contributions of the
Institute for Astronomy, the University of Hawaii, the Pan-STARRS
Project Office, the Max-Planck Society and its participating
institutes, the Max Planck Institute for Astronomy, Heidelberg and the
Max Planck Institute for Extraterrestrial Physics, Garching, The Johns
Hopkins University, Durham University, the University of Edinburgh,
Queen's University Belfast, the Harvard-Smithsonian Center for
Astrophysics, and the Las Cumbres Observatory Global Telescope
Network, Incorporated, the National Central University of Taiwan, and
the National Aeronautics and Space Administration under Grant
No. NNX08AR22G and No. NNX12AR65G issued through the Planetary Science Division of the
NASA Science Mission Directorate.

\clearpage

\appendix
\section{Pan-STARRS1 asteroid database} 
\label{appendix}

Version 1.0 of the Pan-STARRS1 asteroid database is available at
\url{http://www.ifa.hawaii.edu/NEO/}.  It provides derived $H$ and $G$ values for 248,457 asteroids with a
total of 1,242,282 detections spanning the time interval from February
2011 to May 2012 as described in this work.  The 18 column data file
is comma-delimited and each line represents a single asteroid.  The
columns are described in table~\ref{tab.PS1AsteroidDatabaseV1.0}.

\begin{sidewaystable}[htdp]
\tiny
\caption{Pan-STARRS1 asteroid database v1.0 column descriptions.}
\begin{center}
\begin{tabular}{|c|l|l|}
\hline
Col.  &  Col.  &   \\ 
 \#   & value  &  Description \\ 
\hline\hline
1  &  ID  &
  The object's designation in the MPC's 5-character format. The MPC database is accessible online.\footnote{\url{http://www.minorplanetcenter.net/iau/MPCORB/MPCORB.DAT}}\\ 
\hline
2  &  class   &
  The object's taxonomic class as specified by the Sloan Digital Sky Survey \citep{SDSS11} from the Planetary Data System, version 1.1, available online\footnote{\url{http://sbn.psi.edu/pds/resource/sdsstax.html}}. \\ 
   &          & NULL if unknown.\\ 
\hline
3  &  $N$  &
  number of detections used in the fit\\ 
\hline
4  &  $\Delta\alpha$  &
  phase angle range\\ 
\hline
5  &  $H_{B,i}$  &
  initial estimate of the absolute magnitude using the B89 phase curve\\ 
\hline
6  &  $H_{B}$  &
  absolute magnitude derived using our MC technique in the B89 photometric system\\ 
\hline
7  &  $\delta H_{B}$  &
  uncertainty on the absolute magnitude in col. 6\\ 
\hline
8  &  $\Delta H_{B}$  &
  estimated error on the absolute magnitude in col. 6\\ 
\hline
9  &  $H_{M,i}$  &
  initial estimate of the absolute magnitude using the M10 phase curve\\ 
\hline
10 &  $H_{M}$  &
  absolute magnitude derived using our MC technique with the M10 phase curve\\ 
\hline
11 &  $\delta H_{M}$  &
  uncertainty on the absolute magnitude in col. 10\\ 
\hline
12 &  $\Delta H_{M}$  &
  estimated error on the absolute magnitude in col. 10\\ 
\hline
13 &  $G_B$  &
  slope parameter derived using our MC technique in the B89 photometric system\\ 
\hline
14 &  $\delta G_B$  &
  uncertainty on the slope parameter in col. 13\\ 
\hline
15 &  $\Delta G_B$  &
  estimated error on the slope parameter in col. 13\\ 
\hline
16 &  $G_M$  &
  slope parameter derived using our MC technique in the M10 photometric system\\ 
\hline
17 &  $\delta G_M$  &
  uncertainty on the slope parameter in col. 16\\ 
\hline
18 &  $\Delta G_M$  &
  estimated error on the slope parameter in col. 16\\ 
\hline
\end{tabular}
\end{center}
\label{tab.PS1AsteroidDatabaseV1.0}
\end{sidewaystable}


\end{document}